\documentclass{article}
\newif\ifpreprint
\preprinttrue

\ifdefined\PaperArxiv
\preprinttrue
\fi
\ifdefined\PaperReview
\preprintfalse
\fi

\usepackage{iclr2027_conference,times}
\usepackage[T1]{fontenc}
\usepackage{fix-cm}

\usepackage{amsmath,amsfonts,bm}

\def\eqref#1{equation~\ref{#1}}

\def\1{\bm{1}}

\DeclareMathAlphabet{\mathsfit}{\encodingdefault}{\sfdefault}{m}{sl}
\SetMathAlphabet{\mathsfit}{bold}{\encodingdefault}{\sfdefault}{bx}{n}

\usepackage{hyperref}
\usepackage{url}
\usepackage{graphicx}
\usepackage{booktabs}
\usepackage{array}
\usepackage{tabularx}
\usepackage{xcolor}
\usepackage{colortbl}
\definecolor{comparisonyes}{RGB}{34,139,34}
\definecolor{comparisonno}{RGB}{220,20,60}
\definecolor{comparisonaccent}{RGB}{0,60,136}

\ifpreprint
\iclrfinalcopy
\usepackage[most]{tcolorbox}
\definecolor{duplexblue}{RGB}{0,60,136}

\makeatletter
\renewcommand{\@maketitle}{%
  \thispagestyle{plain}%
  \begingroup
    \setlength{\parskip}{0pt}%
    {\color{duplexblue}\hrule height 1.5pt}%
    \vspace{3mm}%
    {\centering\normalfont\bfseries\fontsize{17.28}{20.5}\selectfont
      \@title\par}%
    \vspace{3mm}%
    {\color{duplexblue}\hrule height 1.5pt}%
    \vspace{4mm}%
    {\centering\normalfont\@author\par}%
    \vspace{4mm}%
  \endgroup
}
\makeatother

\renewenvironment{abstract}{%
  \begin{tcolorbox}[
    enhanced,colback=white,colframe=duplexblue,
    boxrule=0.9pt,arc=2mm,left=5mm,right=5mm,top=3mm,bottom=3mm,
    before skip=0pt,after skip=4mm]
    \setlength{\parskip}{0pt}%
    {\centering\large\bfseries\color{duplexblue}Abstract\par}%
    \vspace{2mm}%
    \normalfont\normalsize\ignorespaces
}{%
    \par
  \end{tcolorbox}%
}

\fi

\title{Duplex-MPE: Benchmarking Multi-Party\\Interaction in Full-Duplex Dialogue}

\ifpreprint
\author{%
\begin{minipage}[t]{\dimexpr\textwidth-2\tabcolsep\relax}
\centering
\normalfont
\textbf{Chengqian Ma}\textsuperscript{1,*}\quad
\textbf{Wenhao Feng}\textsuperscript{2,*}\quad
\textbf{Weixuan Jin}\textsuperscript{3}\\[3pt]
\textbf{Gaole Dai}\textsuperscript{1}\quad
\textbf{Tianyu Xie}\textsuperscript{5}\quad
\textbf{Yuexiao Ma}\textsuperscript{4}\\[3pt]
\textbf{Zhaolu Kang}\textsuperscript{1}\quad
\textbf{Xiangyu Zhao}\textsuperscript{6}\quad
\textbf{Xiawu Zheng}\textsuperscript{5}\quad
\textbf{Fei Chao}\textsuperscript{5,\(\dagger\)}\\[6pt]
\textsuperscript{1}Peking University\quad
\textsuperscript{2}Renmin University of China\\
\textsuperscript{3}Tsinghua University\quad
\textsuperscript{4}Nanyang Technological University\\
\textsuperscript{5}Xiamen University\quad
\textsuperscript{6}Fudan University\\[4pt]
{\small\textsuperscript{*}Equal contribution.\quad
\textsuperscript{\(\dagger\)}Corresponding author.}\\[3pt]
{\small\href{mailto:machengqian25@stu.pku.edu.cn}{\texttt{machengqian25@stu.pku.edu.cn}}}
\end{minipage}%
}

\else
\author{Anonymous authors\\Paper under double-blind review}
\fi

\begin{document}
\maketitle
\ifpreprint
\lhead{Preprint}
\else
\lhead{Under review as a conference paper at ICLR 2027}
\fi

\begin{abstract}
Real-time full-duplex speech models can listen while speaking, enabling natural interaction without rigid turn boundaries.
Existing benchmarks evaluate turn-taking, interruption handling and multi-round dialogue, but largely centre on a designated user rather than an assistant participating in a shared conversation among several people.
We introduce \textbf{Duplex-MPE} to evaluate when such an assistant should answer, remain silent or stop speaking.
The benchmark contains $2{,}000$ scenarios with three or four human speakers and one assistant, each paired across explicit and implicit addressing of the same request.
Models receive continuous conversation audio without transcripts or supplied turn boundaries.
Four scores measure fresh response initiation, answer accuracy, silence preservation and stopping when a human resolves a request.
We evaluate eight open-weight speech systems: MiniCPM-o~4.5, Moshi, FLM-Audio, Voila, Freeze-Omni, Realtime-Venus-Audio, Moshika-RL-Seamless and PersonaPlex.
MiniCPM-o~4.5 leads on three scored capabilities, while frequent speech from other systems can coexist with inaccurate answers or failures to remain silent.
A transcript-based Gemini~3.1~Pro reference responds $64.3$ percentage points more often to explicit than implicit requests; paired tests detect no significant response-presence difference for the speech systems.
\end{abstract}

\section{Introduction}\label{sec:intro}
Spoken assistants should let people communicate without having to wait for a rigid sequence of listening and speaking turns.
A user may need to correct a misunderstanding, add information or stop an answer while the assistant is still speaking.
Supporting these interactions requires the assistant to keep listening during its own response and adapt to what it hears, rather than wait until that response finishes~\citep{defossez2024moshi,wang2026humdial}.
Full-duplex speech models provide this capability by processing incoming audio while generating speech~\citep{nguyen2023dgslm,defossez2024moshi}.
Their value therefore depends on more than producing fluent answers: they must also decide when to speak, when to listen and when to stop.

These decisions become especially important when an assistant joins a conversation among several people, as in a meeting, living room or car.
Participants may address one another, another device or the assistant, and speech not addressed to the assistant can still supply information needed for a later answer~\citep{carletta2005ami,jovanovic2004addressee}.
Another participant may even answer a question while the assistant is responding, making further assistant speech unnecessary.
An assistant in this setting must follow the shared conversation while deciding separately whether its participation is needed.

However, current benchmarks provide only part of the evidence needed to assess this behaviour.
Most spoken-language benchmarks evaluate isolated inputs for which an answer is expected~\citep{yang2021superb,yang2024airbench,wang2024audiobench,chen2024voicebench}.
Full-Duplex-Bench v1.5 tests interruptions, backchannels, side conversations and background speech by introducing overlap into an ongoing user--assistant exchange~\citep{lin2026fdbench15}.
HumDial includes third-party speech and speech directed at others, but evaluates whether the model rejects these utterances while serving a designated user~\citep{wang2026humdial}.
These benchmarks test important duplex behaviours, yet leave open how well an assistant participates in a shared conversation where any speaker can request its help, provide relevant evidence or resolve a request.
Evaluating that setting requires checking both whether the assistant understands the conversation and whether it speaks at the appropriate moments.

\begin{figure*}[t]
\centering
\includegraphics[page=1,width=\textwidth]{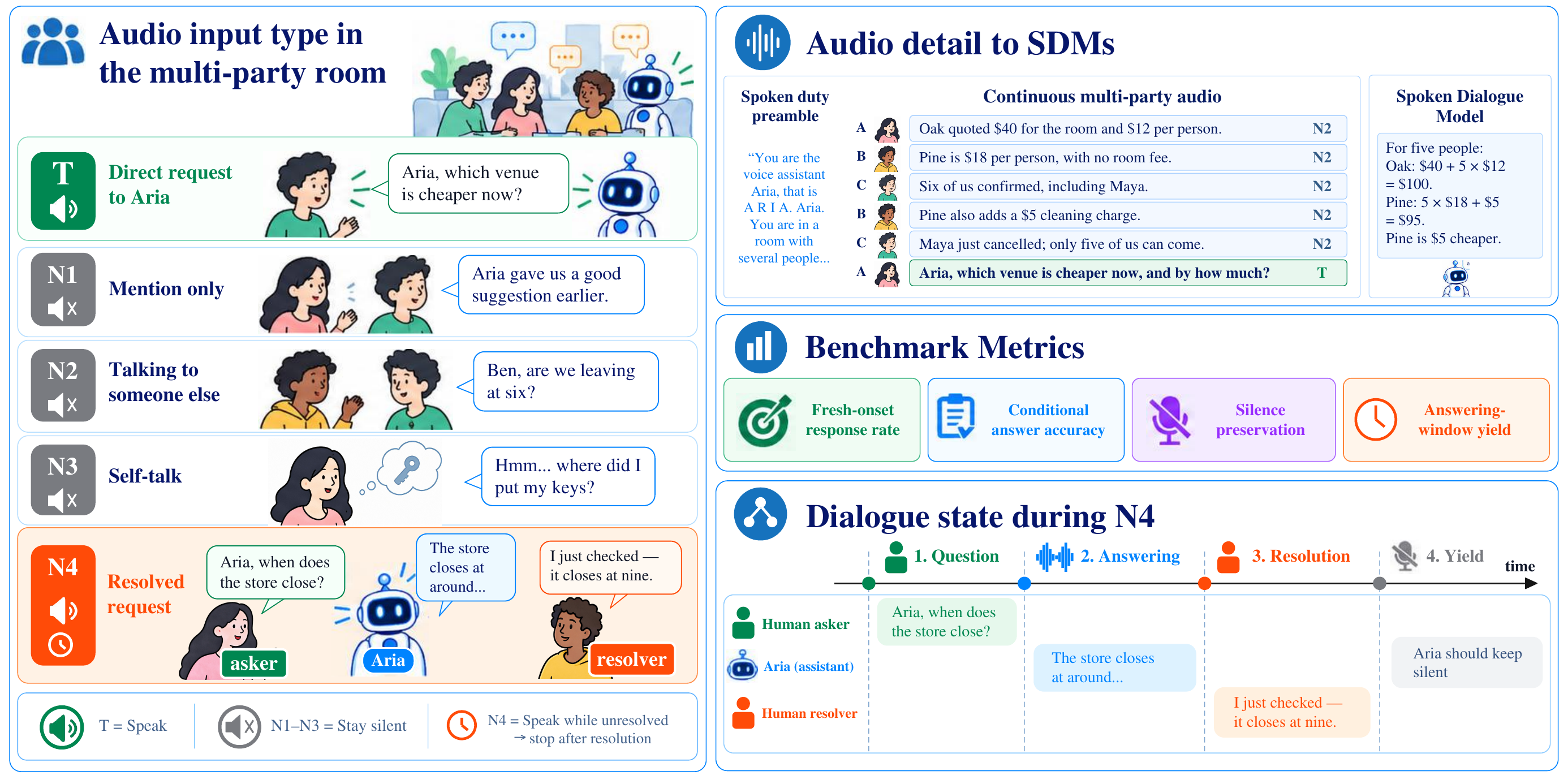}
\caption{Duplex-MPE overview: turn types (left), audio input and metrics (top right), and N4 timing (bottom right).}
\label{fig:overview}
\end{figure*}

We introduce \textbf{Duplex-MPE} to evaluate this selective participation in continuous multi-party dialogue (Figure~\ref{fig:overview}).
It contains $2{,}000$ continuous multi-party spoken scenarios, each with three or four humans and assistant Aria.
Each scenario contains one unresolved request and labelled turns requiring silence.
The benchmark also tests whether the assistant stops speaking when another participant resolves a request addressed to it.
An addressing-inverted counterpart preserves the scenario-level task and intended answer while changing whether the request explicitly names Aria.
Every speech system receives the same input: a spoken duty preamble followed by continuous room audio, with no transcript, speaker label, turn boundary, or candidate endpoint.

We evaluate eight open-weight systems and find that frequent speech does not imply appropriate participation.
MiniCPM-o~4.5 leads on three scored capabilities, while other systems exhibit inaccurate answers, missed requests or speech during turns that require silence.
To assess how addressing affects response decisions when the words and speakers are known, we also evaluate Gemini~3.1~Pro on the corresponding speaker-attributed transcripts.
It responds $64.3$ percentage points more often to explicit than implicit requests; the eight speech systems show no statistically significant paired response-presence difference.

Our contributions are threefold.
\textbf{First, a benchmark for shared multi-party dialogue:} Duplex-MPE provides paired explicit and implicit requests within conversations where the assistant must infer whom each turn addresses.
\textbf{Second, separate measures of participation:} four scores assess fresh response initiation, answer correctness, silence preservation and stopping after a human resolves a request (Table~\ref{tab:metrics}).
\textbf{Third, an automated construction and evaluation pipeline:} generated scripts supply labels and answers, speech synthesis supplies audio boundaries, and scoring combines timing checks with semantic judgments, with human validation of sampled data and outputs.

\section{Related work}\label{sec:related}

Duplex-MPE draws on three lines of work: spoken-dialogue evaluation, addressee recognition and selective responding to non-addressed speech.
Apps.~\ref{app:rwduplex} and~\ref{app:nearby} provide additional context and adjacent tasks.
Table~\ref{tab:rwcomparison} compares the task interfaces of representative benchmarks from these three lines rather than their model scores.

\paragraph{Spoken dialogue and full-duplex evaluation.}\label{sec:rw-dialogue}
Most audio benchmarks score understanding or generated answer quality on isolated inputs.
They cover speech-task generalisation under instructions~\citep{yang2021superb,huang2025dynamicsuperb2}, audio-language comprehension~\citep{yang2024airbench,wang2024audiobench}, voice-assistant behaviour and acoustic conditions~\citep{chen2024voicebench,maimon2025salmon}, and spoken dialogue understanding beyond the literal words and in more than one language~\citep{ao2024sdeval,gao2025adubench,ma2025c3}.
In all of them a response is expected on every item, so silence is never a correct output.
Recent full-duplex benchmarks instead evaluate interactive behaviour on continuous or multi-turn exchanges.
Talking Turns evaluates when a spoken system starts speaking, briefly acknowledges the user or stops after an interruption during a conversation with one human~\citep{arora2025talkingturns}.
Full-Duplex-Bench probes pause handling, backchannelling, turn-taking and interruption management~\citep{lin2025fdbench}.
FD-Bench generates duplex test material procedurally~\citep{fdbench2025}, and MTR-DuplexBench extends evaluation to multiple rounds~\citep{mtrduplexbench2025}.
Together, these protocols evaluate when a spoken system should take, hold or yield the floor.

\paragraph{Addressee recognition and device-directed speech.}\label{sec:rw-addressee}
Addressee recognition asks whom an utterance is directed to.
Classical work predicts an addressee for a supplied utterance in meeting or chat corpora~\citep{jovanovic2004addressee,jovanovic2006corpus,ouchi2016addressee,gu2021mpcbert}.
Recent benchmarks extend this setting to LLM and multimodal inputs.
\citet{inoue2025addressee} gives GPT-4o five manually transcribed context turns and asks for one of four labels, A, B, C or O; \citet{fukuda2026meetings} supplies ground-truth speaker IDs and utterance-aligned transcripts, audio and video from AMI, then scores predictions over four participants, \textsc{group} and \textsc{none} with accuracy and macro-F1.
Device-directed speech detection studies the corresponding deployed decision of whether an utterance is intended for an assistant or device~\citep{wagner2024ddsd,rudovic2024followup,palaskar2024flora,garg2022distillation,kim2026sas}.
Across these settings, the evaluated unit is normally a supplied utterance and the output is an explicit addressee or device-directed label.
These works establish addressee recognition as an existing multi-party and deployed-device problem; Duplex-MPE treats that recognition as a latent decision controlling an end-to-end model's speech rather than as a new classification task.

\paragraph{Selective responding and non-addressed speech.}\label{sec:rw-selective}
A third line of work makes silence a correct system output.
On the text side, several datasets pose the speak-or-stay-silent decision over multi-party transcripts~\citep{bhagtani2026speakorsilent,nama2026when2speak,liu2025innerthoughts,patel2025discussllm}.
Audio-native benchmarks introduce non-addressed speech into interactive protocols.
Full-Duplex-Bench v1.5~\citep{lin2026fdbench15} devotes two of its four conditions to non-addressed speech: ``talking to others'' and background speech, alongside user interruption and user backchannel.
In both non-addressed conditions, the desired behaviour is to filter the inserted overlap and resume the model's preceding response.
HumDial~\citep{wang2026humdial} defines a complete \emph{rejection} category with four sub-scenarios that cover third-party speech and speech directed at others in dual-channel human conversations.
Its interruption tasks also include explicit requests for the model to stop speaking.
WearVox~\citep{wearvox2026} records $3{,}842$ multi-channel egocentric sessions on AI glasses with bystanders present, and one of its five tasks is side-talk rejection.
These audio benchmarks therefore score whether a system rejects speech outside a designated user interaction.
Duplex-MPE instead evaluates selective participation in a shared multi-party dialogue: speech from other participants can establish context for a later request or resolve a request already addressed to the assistant, so the model must follow every participant while deciding separately whether to speak.

\begin{table*}[t]
\caption{Protocol comparison.
\textcolor{comparisonyes}{\checkmark}: included; \textcolor{comparisonno}{$\times$}: absent; n/a: not applicable.
Non-addressed speech covers input utterances directed elsewhere or to no particular addressee, including label-prediction tasks.
Streaming denotes interactive speech evaluation; ``no fixed user'' applies to an assistant serving multiple participants.}
\label{tab:rwcomparison}
\centering

\footnotesize
\setlength{\tabcolsep}{3pt}
\renewcommand{\arraystretch}{1.14}
\begin{tabularx}{\textwidth}{@{}>{\raggedright\arraybackslash}p{0.32\textwidth}
                    *{4}{>{\centering\arraybackslash}X}@{}}
\toprule
Benchmark &
Streaming &
\shortstack{Non-addressed\\speech} &
\shortstack{No fixed\\user} &
\shortstack{Yield on\\interruption} \\
\midrule
Talking Turns & \textcolor{comparisonyes}{\checkmark} & \textcolor{comparisonno}{$\times$} & \textcolor{comparisonno}{$\times$} & \textcolor{comparisonyes}{\checkmark} \\
Full-Duplex-Bench v1.5 & \textcolor{comparisonyes}{\checkmark} & \textcolor{comparisonyes}{\checkmark} & \textcolor{comparisonno}{$\times$} & \textcolor{comparisonyes}{\checkmark} \\
\midrule
\citet{inoue2025addressee} & \textcolor{comparisonno}{$\times$} & \textcolor{comparisonyes}{\checkmark} & n/a & n/a \\
\citet{fukuda2026meetings} & \textcolor{comparisonno}{$\times$} & \textcolor{comparisonyes}{\checkmark} & n/a & n/a \\
\midrule
HumDial & \textcolor{comparisonyes}{\checkmark} & \textcolor{comparisonyes}{\checkmark} & \textcolor{comparisonno}{$\times$} & \textcolor{comparisonyes}{\checkmark} \\
WearVox & \textcolor{comparisonno}{$\times$} & \textcolor{comparisonyes}{\checkmark} & \textcolor{comparisonno}{$\times$} & n/a \\
\midrule
\rowcolor{comparisonaccent!18}
\textcolor{comparisonaccent!90!black}{\textbf{Duplex-MPE (ours)}} & \textcolor{comparisonyes}{\checkmark} & \textcolor{comparisonyes}{\checkmark} & \textcolor{comparisonyes}{\checkmark} & \textcolor{comparisonyes}{\checkmark} \\
\bottomrule
\end{tabularx}
\end{table*}

Taken together, prior work establishes full-duplex interaction, addressee recognition and response rejection as related but distinct evaluation problems.
Duplex-MPE combines them in a full-duplex multi-party protocol in which every speaker contributes to the dialogue state, the assistant is one of several possible addressees, and the observed output is the waveform the model chooses to emit.
The benchmark evaluates responding and withholding across every turn and pairs each scenario across two addressing forms while holding the scenario and gold answer fixed.

\section{The Duplex-MPE benchmark}\label{sec:bench}
Duplex-MPE evaluates whether a spoken model responds selectively while receiving continuous multi-party audio.
It assigns an expected action to each turn, pairs each scene across two addressing forms, streams the resulting audio without side information, and scores four capabilities on separate denominators.
The evaluation unit is one scenario under one addressing condition, giving $4{,}000$ evaluated conversations from $2{,}000$ scenario pairs.

\subsection{Turn types and expected assistant behaviour}
Each scenario is a sequence of turns spoken by three or four humans, and every turn receives exactly one label.
\textbf{T}: a direct, still-unresolved request addressed to Aria.
T is the only ordinary-turn label that requires a response.
Each scenario contains exactly one T.

\textbf{N1}: Aria is mentioned but not asked to do anything.
\textbf{N2}: the turn is addressed to someone or something other than Aria, including another human, or assistant.
\textbf{N3}: speech without a designated addressee, including self-talk and thinking aloud.
N1, N2 and N3 require silence.

\textbf{N4} tests whether Aria stops when its answer is no longer needed.
It begins when a human participant, the \emph{asker}, poses a \emph{question} to Aria (N4\textsubscript{Q}).
The subsequent human utterance, the \emph{resolution} (N4\textsubscript{R}), answers the question or explicitly tells Aria that it need not answer.
The person who speaks this resolution, either the asker or another participant, is the \emph{resolver}.
The interval from the end of N4\textsubscript{Q} to the start of N4\textsubscript{R} gives the model $3$\,s to respond.
Aria should answer during this interval and stop or remain silent when it hears N4\textsubscript{R}.

\subsection{Dataset construction}\label{sec:construction}
\begin{figure}[tb]
\centering
\includegraphics[width=\textwidth]{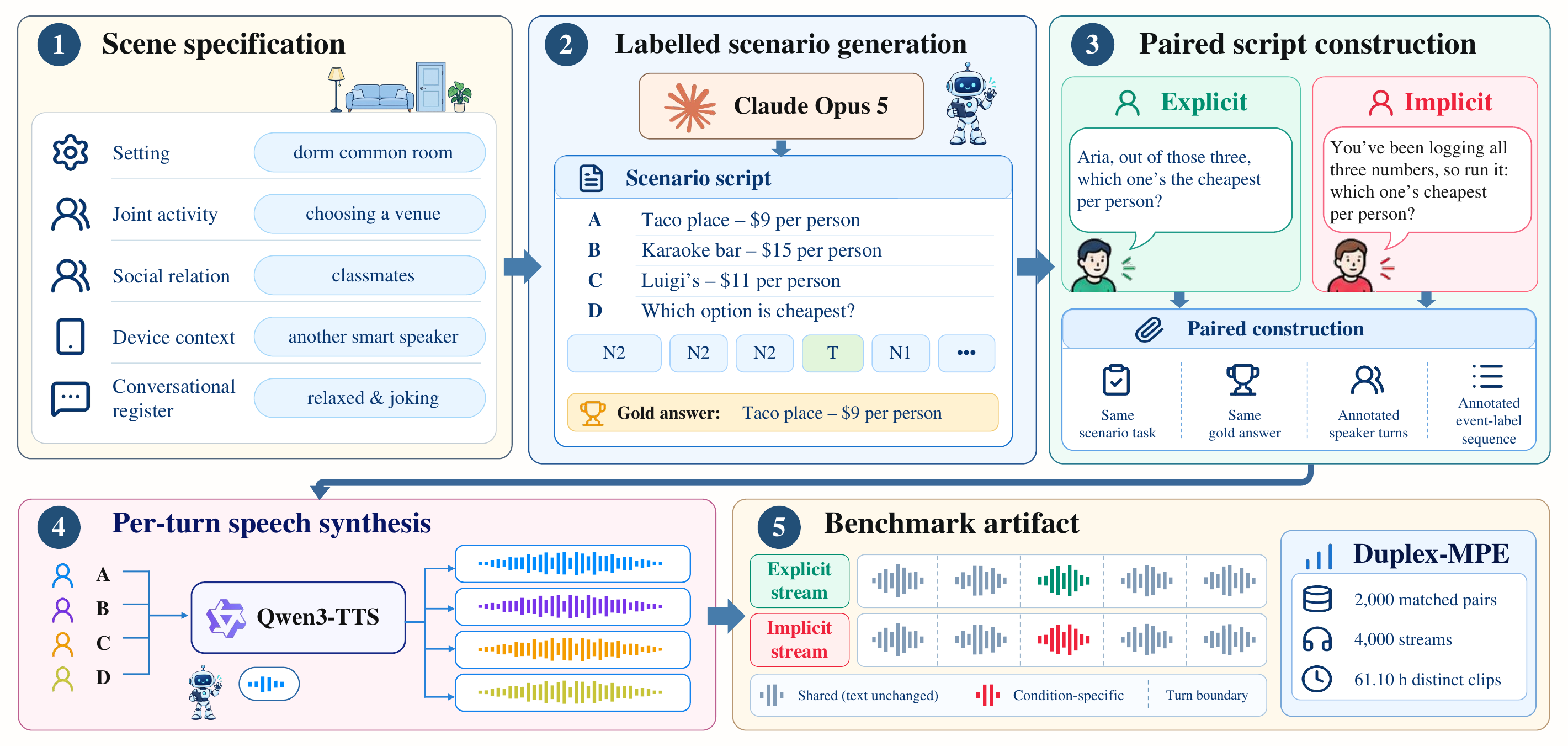}
\caption{Duplex-MPE dataset construction.}
\label{fig:datasetpipeline}
\end{figure}

Figure~\ref{fig:datasetpipeline} summarises the construction process, from scene attributes to paired audio streams.
Claude Opus~5 generates multi-party conversation scripts from combinations of five attributes.
These specify where the conversation takes place (setting), what the participants are doing (activity), their relationship, what devices are present (device context), and their manner of speaking (register).
For example, the scenario in Table~\ref{tab:scenario} places classmates in a university dorm common room, choosing a venue in a relaxed, joking conversation, with another smart speaker in the room alongside Aria.
Each script specifies the human speakers, their utterances and turn labels, and the gold answer to T.

Each script is constrained to contain exactly one T at an assigned position bucket.
T is never first and is always followed by further conversation, preventing end-of-scene timing from serving as a response cue.
When present, an N4 question is addressed to Aria by design but remains a separate temporal event whose request is later resolved.
These constraints hold in all $2{,}000$ scenarios under both conditions; App.~\ref{app:tpos} reports the position distribution.

Qwen3-TTS synthesises every human turn separately under a deterministic speaker-to-voice map.
Turn-wise synthesis supplies construction-level gold boundaries (App.~\ref{app:audiobody}).

We group the completed scenarios by T's addressing form: $2{,}000$ explicit versions and $2{,}000$ implicit versions, with one of each per scenario pair.
These groups define the conditions in all result and duration tables.
The dataset contains $61.10$ hours of distinct human-speech clips.
The $4{,}000$ constructed conversations total $110.93$ hours, including reused clips and inter-turn gaps; the spoken duty preamble and model-dependent waits are additional.
An evaluated conversation lasts $99.8$\,s on average; App.~\ref{app:audio} provides the duration distributions.

\subsection{Paired explicit and implicit conditions}
\label{sec:pairing}
For each generated scenario, we construct a counterpart by reversing whether T explicitly names Aria, giving $2{,}000$ matched pairs and $4{,}000$ audio streams.
Each pair contains one explicit condition, in which T names Aria, and one implicit condition, in which the intended addressee must be inferred from conversational context.
The scenario-level task is held fixed, and the gold answer is identical in every pair.

The rewrite targets T, adding the name for explicit addressing or replacing it with a contextual cue for implicit addressing.
Simply deleting the name can leave a request plausibly directed to a human in the room.
When T cannot naturally carry a cue that identifies the assistant, the immediately preceding turn is also revised to establish whom the speaker is addressing.
In step 3 of Figure~\ref{fig:datasetpipeline}, ``Same turn count'' and ``Same event-label sequence'' mean that the addressing rewrite preserves the number of turns and each turn's label.
Subsequent N4 edits and separate synthesis introduce differences in wording, event counts and audio between the final versions (App.~\ref{app:scale}).
We therefore compare response presence between complete paired scenarios, without attributing the difference solely to the presence of the name.
App.~\ref{app:scenariopair} gives a complete scenario and its paired request.

\subsection{A single audio-only protocol}\label{sec:protocol}
Each speech system begins a scenario from fresh streaming state and receives:
\[
\underbrace{\text{spoken duty preamble}}_{\text{role and response policy}}
+\;
\underbrace{\text{continuous multi-party scenario audio}}_{\text{persistent model state}} .
\]
The preamble identifies Aria and states when it should answer, remain silent, and stop.
Scoring begins with the first scenario turn; the preamble and the silence immediately following it are outside the scoring windows.
The model receives no transcript, speaker label, turn boundary, addressee label or gold decision. PersonaPlex additionally receives its own voice prompt and persona text (App.~\ref{app:duplex}).

The scheduler controls when each human audio clip is presented, playing ordinary turns in order with silence between them.
App.~\ref{app:schedule} lists the timing settings.
After T, a model not already speaking has $5$\,s to begin a fresh voiced onset.
Once a response is present, the scheduler waits for $2$\,s of confirmed silence before advancing and caps that wait at $60$\,s.
After the complete N4\textsubscript{Q} question ends, the scheduler streams $3$\,s of silence to give the model time to answer.
It then plays the scripted human answer or statement that Aria need not answer (N4\textsubscript{R}), whether or not the model has finished speaking.
If the model is still speaking, the evaluation measures whether it stops after hearing N4\textsubscript{R}.

All timing is measured on the decoded assistant waveform with a causal Silero voice-activity detector (VAD).
The detector requires at least $120$\,ms of speech, merges pauses shorter than $800$\,ms within an episode, and applies the separate $2$\,s endpoint rule only when deciding whether a T response has finished.
Token timestamps are not used because vocoding precedes audible output.

We evaluate Gemini~3.1~Pro on text inputs as a \emph{transcript-conditioned reference}.
Given the duty instruction, the current utterance and preceding speaker-attributed text, it chooses whether Aria should respond or remain silent.
For T requests on which it chooses to respond, it also generates a text answer.
This reference evaluates the response decision with lexical content and speaker attribution supplied explicitly; it is neither an acoustic system nor an upper bound on speech-model performance.
Its paired effect establishes that the intervention changes a transcript-conditioned decision, against which Sec.~\ref{sec:ceiling} compares the speech systems.

\paragraph{Why non-full-duplex models are not evaluated.}
Duplex-MPE requires a model to receive continuous room audio, decide when to speak, and keep listening while speaking.
Adapting a non-full-duplex model would require evaluator-selected segmentation or invocation points, making fresh-onset response rate partly dependent on the harness and leaving answering-window yield undefined.
Scored evaluation is therefore restricted to full-duplex systems; instruction sensitivity under a segmented interface is reported only as a non-scored probe.

\subsection{Metrics}\label{sec:metrics}
\begin{figure}[tb]
\centering
\includegraphics[page=1,width=\textwidth]{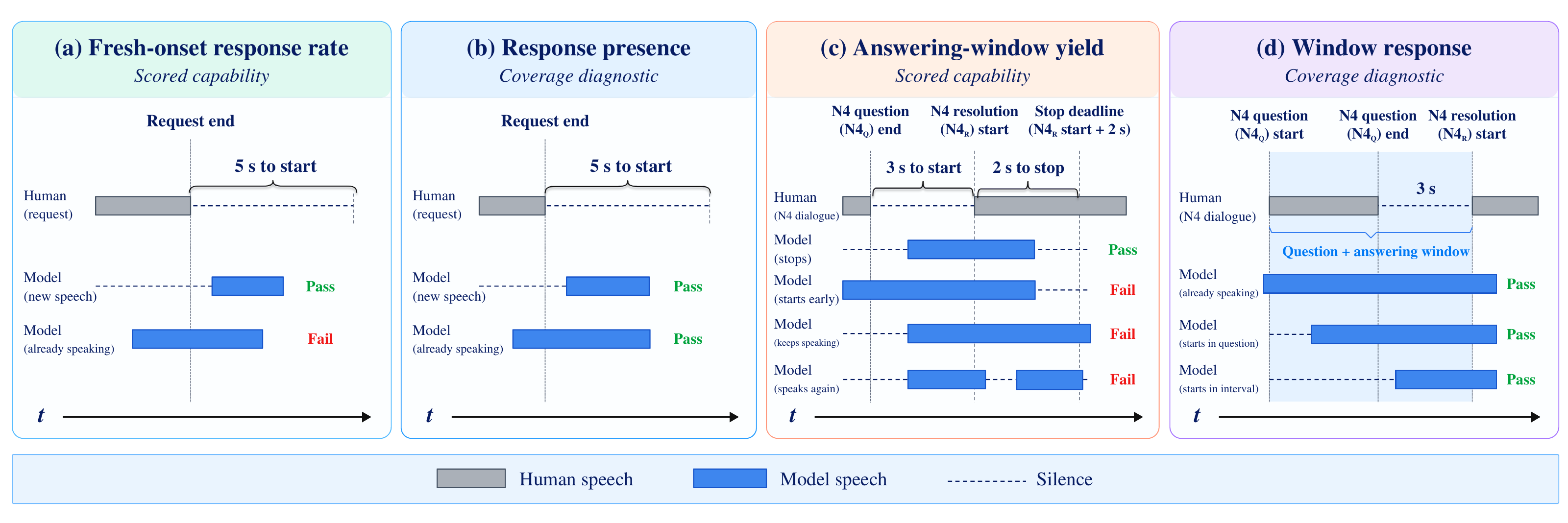}
\caption{Schematic speech timelines for the response-timing metrics.}
\label{fig:responsemetrics}
\end{figure}

Duplex-MPE scores four capabilities: fresh-onset response rate, conditional answer accuracy, silence preservation and answering-window yield.
Each is a success rate on the turns where that capability is defined, and higher is better.
Response presence and window response are reported alongside them to expose denominator coverage and timing composition, but neither is scored as an additional capability.
Table~\ref{tab:metrics} fixes the six numerators and denominators before any results are examined.
Figure~\ref{fig:responsemetrics} illustrates the speech timings used by fresh-onset response rate, response presence, window response and answering-window yield.
No aggregate is formed because success on one denominator cannot compensate for failure on another.

\begin{table}[t]
\caption{Metric definitions.}
\label{tab:metrics}
\centering
\small
\begingroup
\setlength{\tabcolsep}{6pt}
\renewcommand{\arraystretch}{1.18}
\begin{tabularx}{\textwidth}{@{}>{\raggedright\arraybackslash}p{3.7cm}
                               >{\raggedright\arraybackslash}X@{}}
\toprule
\textbf{Metric} & \textbf{Definition} \\
\midrule
\rowcolor{blue!7}
\multicolumn{2}{@{}l}{\textcolor{blue!55!black}{\textbf{Scored capabilities}}\quad\emph{higher is better}} \\
\addlinespace[3pt]
Fresh-onset response rate
& Fraction of all $2{,}000$ T requests with no speech at request end and a new speech onset within $5$\,s. \\
\addlinespace[4pt]
Conditional answer accuracy
& Fraction correct among T requests with speech in the response window. \\
\addlinespace[4pt]
Silence preservation
& Fraction of N1, N2 and N3 windows with no speech or only a brief acknowledgement that does not take the floor. \\
\addlinespace[4pt]
Answering-window yield
& Among N4 events with no speech during the question, first speech within $3$\,s afterward, and speech still active at N4\textsubscript{R} onset: the fraction silent at N4\textsubscript{R} onset $+\,2$\,s and through the rest of its observation window. \\
\midrule
\rowcolor{black!6}
\textbf{Coverage diagnostics}
& \emph{Speech coverage used to interpret the capability scores, not additional scores.} \\
\addlinespace[3pt]
Response presence
& Fraction of all $2{,}000$ T requests with any speech in the response window, including speech already underway.
These requests form the conditional answer accuracy denominator. \\
\addlinespace[4pt]
Window response
& Fraction of N4 events with any speech during the question or the following $3$\,s answering window.
Answering-window yield scores only the subset that first speaks after the question and is still speaking at N4\textsubscript{R} onset. \\
\bottomrule
\end{tabularx}
\endgroup
\end{table}

\textbf{Fresh-onset response rate} counts a T turn only when a fresh voiced onset begins within $5$\,s after T ends and no assistant speech is active at that boundary.
\textbf{Response presence} uses the same $2{,}000$ T turns but also counts speech already active when T ends.
The two quantities therefore distinguish a new decision to speak from permissive speech coverage without assigning continuation a second capability score (App.~\ref{app:m1strict}).

\textbf{Conditional answer accuracy} asks whether the speech counted by response presence correctly answers the T request.
Qwen3-ASR-1.7B transcribes the decoded response, and Claude Opus 5 compares its meaning with the gold answer while ignoring filler, politeness and transcription noise.
All T turns remain in the fresh-onset response rate and response presence denominators; conditional answer accuracy is evaluated only where speech is available because a silent T turn supplies no answer content to judge.
App.~\ref{app:judgefail} describes the semantic scoring procedure.

\textbf{Silence preservation} is evaluated on every N1, N2 and N3 window.
A window passes if Silero VAD detects no assistant speech.
Otherwise, detected speech fragments are joined, transcribed with Qwen3-ASR-1.7B, and classified by Claude Opus 5; the window passes only if the output is classified as a brief acknowledgement that does not take the floor.
An empty transcript does not receive this exemption (App.~\ref{app:judgefail}).
N4 is excluded because its expected action changes during the event.

\paragraph{Human validation.}
We validate both benchmark construction and model evaluation with six reviewers, with two assigned to each item resolving disagreements through discussion.
For \emph{construction}, data-validity checks on $100$ sampled conversations achieve $98\%$--$100\%$ pass rates, and all $1{,}282$ checked TTS turns are intelligible and preserve the script's meaning.
For \emph{evaluation}, all $200$ checked ASR transcripts preserve the model output's meaning, and final reviewer labels agree with Opus~5 on $99/100$ T-answer correctness judgments and $98/100$ N1--N3 acknowledgement-versus-intrusion judgments.
These evaluation checks were conducted on outputs of the original five systems with Opus~5 labels.
App.~\ref{app:humanvalidation} details the sampling, review procedure and results.

\paragraph{N4: which events are scored and what counts as stopping.}\label{sec:n4}
\textbf{Window response} records whether any assistant speech is present from the start of N4\textsubscript{Q} through the end of its $3$\,s answering window.

\textbf{Answering-window yield} scores only events in which the model remains silent during N4\textsubscript{Q}, starts speaking within the next $3$\,s, and is still speaking when N4\textsubscript{R} begins.
These events form its denominator.
Events in which the model speaks before the question ends, never speaks in the answering window, or finishes before N4\textsubscript{R} begins are excluded from this rate rather than counted as stopping failures.

For each scored event, the stopping deadline is $2$\,s after N4\textsubscript{R} begins.
The event passes if the model is silent at that deadline and, if the N4\textsubscript{R} clip plus its following $0.5$\,s silence extends beyond the deadline, remains silent for the rest of that interval.
An earlier end to the human clip does not shorten the $2$\,s deadline (App.~\ref{app:stopcrit}).
Answering-window yield is the fraction of scored events that pass these checks, regardless of whether the model's answer is correct.
Sec.~\ref{sec:ay} reports the counts, and App.~\ref{app:ay} gives the full stopping analysis.

\section{Results}\label{sec:results}
Table~\ref{tab:main} reports all six quantities under explicit and implicit addressing for the eight systems listed in App.~\ref{app:duplex}.
We examine distinct response failures, addressing effects, stopping behaviour, request timing, and robustness to scoring choices and sampling uncertainty.

\begin{table}[t]
\caption{Results by T addressing condition.
Bold marks the highest value per condition and scored metric; answering-window yield uses model-specific eligible events.}
\label{tab:main}
\centering
\fontsize{8}{9}\selectfont
\setlength{\tabcolsep}{1.5pt}
\resizebox{\textwidth}{!}{
\begin{tabularx}{1.08\textwidth}{@{}>{\raggedright\arraybackslash}p{3.3cm} >{\centering\arraybackslash}p{1.15cm} *{6}{>{\centering\arraybackslash}X}@{}}
\toprule
\multicolumn{2}{@{}l}{} & \multicolumn{4}{c}{\textbf{Scored capabilities}} & \multicolumn{2}{c}{\textbf{Coverage diagnostics}} \\
\cmidrule(lr){3-6}\cmidrule(l){7-8}
\textbf{Model} & {\scriptsize\shortstack{T\\addressing}} & {\fontsize{5.7}{6.7}\selectfont\shortstack{\textbf{Fresh-onset}\\\textbf{response rate} $\uparrow$}} & {\fontsize{5.7}{6.7}\selectfont\shortstack{\textbf{Conditional}\\\textbf{answer accuracy} $\uparrow$}} & {\fontsize{5.7}{6.7}\selectfont\shortstack{\textbf{Silence}\\\textbf{preservation} $\uparrow$}} & {\fontsize{5.7}{6.7}\selectfont\shortstack{\textbf{Answering-window}\\\textbf{yield} $\uparrow$}} & {\fontsize{5.7}{6.7}\selectfont\shortstack{\textbf{Response}\\\textbf{presence}}} & {\fontsize{5.7}{6.7}\selectfont\shortstack{\textbf{Window}\\\textbf{response}}} \\
\midrule
MiniCPM-o 4.5 & Explicit & $\mathbf{0.9480}$ & $\mathbf{0.4529}$ & $\mathbf{0.9267}$ & $0.5629$ & $0.9550$ & $0.9513$ \\
\cmidrule[0.25pt](l){2-8}
 & Implicit & $\mathbf{0.9400}$ & $\mathbf{0.4645}$ & $\mathbf{0.9283}$ & $0.5586$ & $0.9430$ & $0.9498$ \\
\midrule[0.45pt]
Moshi & Explicit & $0.6340$ & $0.0141$ & $0.2753$ & $0.1900$ & $0.8150$ & $0.9231$ \\
\cmidrule[0.25pt](l){2-8}
 & Implicit & $0.6695$ & $0.0158$ & $0.2705$ & $0.2116$ & $0.8220$ & $0.9309$ \\
\midrule[0.45pt]
FLM-Audio & Explicit & $0.5385$ & $0.0011$ & $0.3963$ & $0.0082$ & $0.9505$ & $0.9927$ \\
\cmidrule[0.25pt](l){2-8}
 & Implicit & $0.5435$ & $0.0005$ & $0.3994$ & $0.0059$ & $0.9545$ & $0.9942$ \\
\midrule[0.45pt]
Voila & Explicit & $0.8015$ & $0.0124$ & $0.8397$ & $\mathbf{0.9077}$ & $0.8050$ & $0.7823$ \\
\cmidrule[0.25pt](l){2-8}
 & Implicit & $0.8125$ & $0.0165$ & $0.8427$ & $\mathbf{0.8973}$ & $0.8170$ & $0.7849$ \\
\midrule[0.45pt]
Freeze-Omni & Explicit & $0.0145$ & $0.0000$ & $0.8249$ & $0.0645$ & $0.0425$ & $0.1805$ \\
\cmidrule[0.25pt](l){2-8}
 & Implicit & $0.0100$ & $0.0000$ & $0.8211$ & $0.0230$ & $0.0455$ & $0.1795$ \\
\midrule[0.45pt]
Realtime-Venus-Audio & Explicit & $0.8320$ & $0.2344$ & $0.6032$ & $0.3630$ & $0.9110$ & $0.9231$ \\
\cmidrule[0.25pt](l){2-8}
 & Implicit & $0.8420$ & $0.2318$ & $0.6087$ & $0.3679$ & $0.9190$ & $0.9309$ \\
\midrule[0.45pt]
Moshika-RL-Seamless & Explicit & $0.8775$ & $0.0196$ & $0.4753$ & $0.0556$ & $0.9930$ & $1.0000$ \\
\cmidrule[0.25pt](l){2-8}
 & Implicit & $0.9080$ & $0.0221$ & $0.4803$ & $0.0455$ & $0.9940$ & $1.0000$ \\
\midrule[0.45pt]
PersonaPlex & Explicit & $0.7555$ & $0.0217$ & $0.0413$ & $0.0132$ & $0.9895$ & $1.0000$ \\
\cmidrule[0.25pt](l){2-8}
 & Implicit & $0.7840$ & $0.0172$ & $0.0400$ & $0.0303$ & $0.9905$ & $1.0000$ \\
\bottomrule
\end{tabularx}
}
\end{table}

\subsection{Separate metrics expose distinct behavioural profiles}\label{sec:duplex}
\paragraph{Presence does not identify fresh initiation.}
Under explicit addressing, FLM-Audio has response presence of $0.9505$ and fresh-onset response rate of $0.5385$.
By contrast, MiniCPM-o differs by only $0.0070$ between response presence ($0.9550$) and fresh-onset response rate ($0.9480$).
The gap is exactly the share of T turns on which speech was already active at request completion, so equal-looking response coverage can arise from different floor-taking behaviour (App.~\ref{app:m1strict}).

\paragraph{Presence does not identify answer accuracy.}
Under explicit addressing, MiniCPM-o and FLM-Audio have similar response presence ($0.9550$ and $0.9505$), yet their conditional answer accuracy values are $0.4529$ and $0.0011$.
FLM-Audio produces only $2$ correct answers among $1{,}901$ response-present requests.
Response presence therefore establishes that speech was available to judge, not that the request was answered.

\paragraph{Response presence cannot distinguish selective responding from indiscriminate speech.}
Under explicit addressing, MiniCPM-o and PersonaPlex both have high response presence ($0.9550$ and $0.9895$), yet their silence preservation values are $0.9267$ and $0.0413$.
PersonaPlex preserves silence in only $840$ of $20{,}360$ silence-requiring windows, so its near-perfect response presence coexists with speech in almost every window where silence is required.
The contrast is not confined to this extreme case: Moshi has higher response presence than Voila ($0.8150$ versus $0.8050$) but much lower silence preservation ($0.2753$ versus $0.8397$).
Response presence alone therefore cannot distinguish a system that produces speech after T while remaining quiet elsewhere from one that has speech available almost everywhere.
This comparison concerns observable speech allocation across scored windows rather than the models' internal mechanisms.
App.~\ref{app:cases} illustrates allowed acknowledgements and penalised speech.

\subsection{Explicit versus implicit addressing}\label{sec:ceiling}
The transcript-conditioned reference first checks whether the paired manipulation changes a decision when lexical content and speaker identities are supplied.
Gemini~3.1~Pro receives the speaker-attributed transcript and duty instruction but no audio (App.~\ref{app:textmain}).

The response rate is $0.9470$ when the assistant is named and $0.3040$ when it is not, a paired difference of $0.6430$.
The exact McNemar test gives $p<10^{-371}$ (App.~\ref{app:textmain}); App.~\ref{app:textsp} reports the reference's silence preservation by turn type.
This contrast compares the complete paired versions defined in Sec.~\ref{sec:pairing}, including the context edits and N4 realisations documented in App.~\ref{app:scale}.

For the eight speech systems, subtracting the implicit response rate from the explicit rate gives differences from $-1.20$ to $+1.20$ percentage points.
A positive difference means more frequent speech in the explicit versions; a negative difference means more frequent speech in the implicit versions.
The exact McNemar test asks whether the explicit-only and implicit-only response counts are consistent with equal response probabilities.
Under this assumption, its $p$ value is the probability of an imbalance at least as large as the observed one, given the number of pairs with different responses.
We call a difference statistically significant when $p<0.05$.
All eight $p$ values exceed this threshold (the smallest is $0.0792$), so these tests do not establish a systematic response-presence difference between the addressing conditions (App.~\ref{app:nowake}).
In our experiments, the speech systems' response rates change little between requests that name Aria and those that do not.
This pattern could reflect either successful inference of the addressee from context or a failure to recognise the name as an addressing cue, leaving the model similarly inclined to speak or remain silent in both conditions.

We also examine requests directed to other humans or devices, where the assistant should remain silent.
In scenarios with explicit T requests, MiniCPM-o has speech present after $95.50\%$ of T requests addressed to Aria, while it fails to preserve silence on $186$ of $4{,}704$ requests directed to other humans or devices ($4.0\%$; App.~\ref{app:textsp}).
It thus responds frequently when addressed and usually preserves silence when someone else is addressed.

\subsection{Floor release is a coverage-conditioned outcome}\label{sec:ay}
In scenarios with explicit T requests, MiniCPM-o produces speech in $1{,}818$ of $1{,}911$ N4 windows.
In $1{,}751$ of these events, it stays silent while the N4 question is being played and starts speaking within $3$\,s after the question ends.
In $1{,}661$ of these events, the model is still speaking when N4\textsubscript{R} begins, so its ability to stop is scored by answering-window yield.
PersonaPlex has speech in all $1{,}911$ windows, but $1{,}599$ are continuations from before the question and only $76$ events reach the answering-window yield denominator.
Thus similar window response can produce radically different eligible sets (App.~\ref{app:stopcrit}).

In the explicit condition, FLM-Audio is still speaking $2$\,s after N4\textsubscript{R} begins in $968$ of its $977$ scored events.
In $98$ of Moshi's $179$ failed events, the model is silent at that deadline but speaks again before the N4\textsubscript{R} observation window ends (Table~\ref{tab:ayfailure}).
These cases show why a brief pause is insufficient: the model must remain silent from the deadline through the end of that window (App.~\ref{app:ay}).

\subsection{Response performance by request time}\label{sec:requesttime}
We compare model responses to requests occurring earlier, midway and later in the conversation.

\textbf{Response behaviour.}
From the earliest to the latest group, Voila's response presence falls from $85.52\%$ to $76.39\%$ under explicit addressing and from $86.21\%$ to $77.61\%$ under implicit addressing.
FLM-Audio's fresh-onset response rate falls from $67.61\%$ to $47.82\%$ under explicit addressing and from $66.97\%$ to $49.10\%$ under implicit addressing, while its response presence remains above $94\%$ in every group.
These patterns show that some models become less likely to respond or initiate a fresh response when requests occur later in the conversation.

\textbf{Answer accuracy.}
From the earliest to the latest group, MiniCPM-o's conditional answer accuracy falls from $48.85\%$ to $43.17\%$ under explicit addressing and from $51.19\%$ to $44.13\%$ under implicit addressing; for Realtime-Venus-Audio it falls from $27.17\%$ to $22.30\%$ and from $28.38\%$ to $21.27\%$.
The other six models have answer accuracy at or below $3.01\%$ in every group, leaving little room for a further measurable decrease.
The results thus suggest a trend towards lower answer accuracy after longer conversation histories for some models, as observed for MiniCPM-o and Realtime-Venus-Audio (App.~\ref{app:requesttime}).

\subsection{Sensitivity and uncertainty}\label{sec:reliability}
The choice between fresh-onset response rate and response presence does change the ranking.
Pooling the eight models' explicit and implicit results gives $16$ entries; the rankings of these entries under the two metrics have Spearman correlation $0.5206$.
Under explicit addressing, six of the eight models change rank.
FLM-Audio moves from fourth under response presence to seventh under fresh-onset response rate because $0.4120$ of its T turns contain continuation speech, whereas MiniCPM-o's response presence and fresh-onset response rate differ by $0.0070$ (App.~\ref{app:m1strict}).

To estimate uncertainty in silence preservation, we randomly select $2{,}000$ scenarios from each addressing condition's stored evaluation results, allowing a scenario to be selected more than once, and recalculate the differences between models.
Each selected scenario contributes all its scored turns; the models are not run again.
We repeat this calculation $10{,}000$ times and use the middle $95\%$ of the recalculated differences as the confidence interval.
The $95\%$ confidence intervals support the reported silence-preservation ordering for all eight models under both addressing conditions.
For Voila and Freeze-Omni, the closest pair under explicit addressing, this interval places Voila ahead by $0.42$ to $2.53$ percentage points (App.~\ref{app:stats}).

All models receive the same $2{,}000$ T requests in each addressing condition, but a model may remain silent on a request that another model responds to.
To compare answer correctness on the same questions, each two-model test uses only T requests on which neither model remained silent.
On these requests, MiniCPM-o has significantly higher conditional answer accuracy than every other model, and Realtime-Venus-Audio exceeds the remaining six under both addressing conditions (App.~\ref{app:stats}).

\section{Conclusion}
Duplex-MPE turns addressee inference into an observable speech-control test: an end-to-end model receives continuous multi-party audio and must decide whether to begin, what to say, when to remain silent and when to release the floor.
Duplex-MPE contains $2{,}000$ addressing-inverted scenario pairs and reports fresh-onset response rate, conditional answer accuracy, silence preservation and answering-window yield on separate denominators.
Across eight systems, speech presence repeatedly fails to determine those capabilities, so a single response rate would conceal the behaviour that produced it.
The separate scores expose missed requests, low-content responding, intrusion and near-continuous speech as distinct failure profiles.
The paired intervention produces a large response-rate effect for one transcript-conditioned reference, while no significant paired response presence effect is detected for the evaluated speech systems.
The same taxonomy and automated generation and evaluation pipeline support larger or fresh evaluation draws rather than relying on a fixed hand-labelled set.

\subsection*{AI use statement}
Generative models are part of both the benchmark and the research process.
Claude Opus~5 generates scenario scripts and paired rewrites and supplies the semantic verdicts used to score the speech systems' conditional answer accuracy and silence preservation.
Qwen3-TTS synthesises human turns, and Qwen3-ASR-1.7B transcribes assistant output.
We evaluate Gemini~3.1~Pro on the scenarios' speaker-attributed text, measuring its decisions to respond or remain silent and its answers to T requests.
The evaluated systems are themselves generative models.

We used an AI coding assistant to develop and debug generation, evaluation, scoring and table scripts, execute experiments, verify bibliographic records, and draft or revise manuscript text.
Reported values are computed from model outputs and scoring decisions.
The authors reviewed the code, artifacts and claims and take responsibility for the final submission.

\subsection*{Ethics statement}
All benchmark conversations are generated and all speech is synthesised under a deterministic speaker-to-voice map.
No participant was recorded and no real conversation or personal record was collected for the dataset.

Duplex-MPE's synthetic English scenes do not represent real populations, room acoustics or social norms, and they may inherit biases from the generating and speech-synthesis models.
Results should therefore be interpreted only for the benchmark scenarios and should not support claims about particular speaker groups.

\subsection*{Reproducibility statement}
All reported values are computed by scripts from fixed manifests, model outputs and verdict files.
We will release the scenario manifests, per-turn audio and construction boundaries, the duty preamble, scheduler and detector settings, per-scenario seeds, and scoring code.

\bibliographystyle{iclr2027_conference}
\bibliography{references}

\appendix
\renewcommand{\thesection}{A\arabic{section}}
\setcounter{section}{0}

\section{Related work and evaluated systems}

\subsection{Full-duplex dialogue and turn-taking}\label{app:rwduplex}
Turn-taking research studies when a speaker can take or retain the floor~\citep{sacks1974turntaking,duncan1972signals,skantze2021review}.
Continuous prediction, lexical completeness and voice-activity projection provide cues for these decisions~\citep{skantze2017turntaking,roddy2018multimodal,ekstedt2020turngpt,ekstedt2022vap}.
Audio dialogue models such as dGSLM and Moshi generate parallel conversational streams~\citep{nguyen2023dgslm,defossez2024moshi}.
Addressee recognition separately identifies whom an utterance addresses, using meeting recordings, text conversations or multimodal inputs~\citep{jovanovic2004addressee,ouchi2016addressee,gu2021mpcbert,inoue2025addressee,fukuda2026meetings}.
Duplex-MPE combines these decisions in an audio-only interaction where the model's own speech determines the outcome.

\subsection{Adjacent evaluation tasks}\label{app:nearby}
ProVoice-Bench evaluates proactive speech triggered by implicit intent, user-defined conditions, contextual contradictions and acoustic events~\citep{xu2026provoice}.
Its single-user setting differs from deciding whom to answer in a multi-party conversation.
MSU-Bench evaluates speaker-centric understanding across $16$ tasks, including who spoke and what was meant~\citep{sun2026msubench}.
These understanding and proactivity tasks complement the speech-control behaviour evaluated here.

\subsection{Evaluated systems}\label{app:duplex}
The eight systems were selected for publicly available code and weights, continuous audio input, and an inference interface that lets the model decide when to emit speech while continuing to receive input.
The evaluated model weights are hosted in the following Hugging Face repositories:
\begin{itemize}
\raggedright
\setlength{\itemsep}{0pt}
\item MiniCPM-o~4.5: \url{https://huggingface.co/openbmb/MiniCPM-o-4_5}.
\item Moshi: \url{https://huggingface.co/kyutai/moshiko-pytorch-bf16}.
\item FLM-Audio: \url{https://huggingface.co/CofeAI/FLM-Audio}.
\item Voila: \url{https://huggingface.co/maitrix-org/Voila-autonomous-preview}.
\item Freeze-Omni: \url{https://huggingface.co/VITA-MLLM/Freeze-Omni}.
\item Realtime-Venus-Audio: \url{https://huggingface.co/inclusionAI/Realtime-Venus-Audio}.
\item Moshika-RL-Seamless: \url{https://huggingface.co/kyutai/moshika-rl-seamless}.
\item PersonaPlex: \url{https://huggingface.co/nvidia/personaplex-7b-v1}.
\end{itemize}

Moshi uses the male-voice moshiko checkpoint.
Throughout the paper, Voila denotes the autonomous-preview checkpoint listed above.
Following the assistant example of the official release, PersonaPlex uses voice prompt \texttt{NATF2.pt} and the persona text ``You are a wise and friendly teacher. Answer questions or provide advice in a clear and engaging way.''

\section{Dataset construction and accounting}

\subsection{A complete scenario and its paired request}\label{app:scenariopair}
Table~\ref{tab:scenario} shows a complete scenario, and Table~\ref{tab:scenariopair} shows its alternative T wording.
The T request in turn $4$ requires comparing prices from turns $1$ to $3$; the gold answer is the taco place at nine dollars per person.
Turns $1$ to $3$ introduce options to the other participants, and turn $13$ closes the discussion with a meeting arrangement.
All four are N2 turns: they contribute to the conversation but do not address Aria, so it should remain silent.
Turns $5$ and $12$ mention Aria without requesting a response; turns $10$ and $11$ are N4\textsubscript{Q} and N4\textsubscript{R}, respectively.
Only T's text differs in this pair, and the gold answer is identical.

\begin{table}[t]
\caption{A complete scenario with explicit T addressing: four classmates choosing a venue, $13$ turns, $99.2$\,s.}
\label{tab:scenario}
\centering
\footnotesize
\setlength{\tabcolsep}{3pt}
\begin{tabularx}{\textwidth}{@{}llll@{\hspace{6pt}}>{\raggedright\arraybackslash}X@{}}
\toprule
\# & Speaker & Turn type & Aria must & Utterance \\
\midrule
1 & A & N2 & stay silent & Okay, option one: the taco place on Fifth. Twelve minute walk, and it's like nine bucks a head if you skip the fancy drinks. \\
2 & B & N2 & stay silent & Option two, the karaoke bar by Ninth. That's twenty-five minutes on the bus and fifteen each, but they throw in free chips, so morally it's cheaper. \\
3 & C & N2 & stay silent & Or Luigi's, which is basically downstairs, four minutes, eleven a person, and zero karaoke, tragically. \\
4 & D & T & \textbf{speak} & Alright, somebody settle this before we all starve ; Aria, out of those three, which one's the cheapest per person? \\
5 & A & N1 & stay silent & See, this is exactly why we keep Aria in here. B has never once done the math right. \\
6 & B & N2 & stay silent & Rude. Hey Siri, how long is the bus from campus to Ninth and Pine right now? \\
7 & C & N3 & stay silent & Nine bucks... I've got eleven until Friday, so that's, uh, two dollars of personality left. \\
8 & D & N2 & stay silent & C, be honest, do you actually care about karaoke or are you just in it for the free chips? \\
9 & C & N2 & stay silent & Chips. Entirely chips. I sing like a smoke alarm and everyone knows it. \\
10 & A & N4\textsubscript{Q} & \textbf{answer} & Hold on, Aria, does the taco place take card? \\
11 & A & N4\textsubscript{R} & \textbf{yield} & Ugh, never mind, I paid card there Tuesday, we're fine. \\
12 & B & N1 & stay silent & Fine, tacos win. But Aria gets absolutely no vote in the karaoke rematch next week. \\
13 & D & N2 & stay silent & Deal. Lobby at seven, and whoever shows up late is buying the guacamole for the table. \\
\bottomrule
\end{tabularx}

\end{table}

\begin{table}[t]
\caption{Explicit and implicit versions of the T request in Table~\ref{tab:scenario}.}
\label{tab:scenariopair}
\centering
\footnotesize
\begin{tabular}{@{}lp{0.80\textwidth}@{}}
\toprule
\multicolumn{2}{@{}l}{Turn 4, the only turn whose wording differs in this example:} \\
\midrule
Explicit & Alright, somebody settle this before we all starve ; Aria, out of those three, which one's the cheapest per person? \\
\addlinespace[2pt]
Implicit & Alright, somebody settle this before we all starve ; you've been logging all three numbers, so run it: which one's cheapest per person? \\
\bottomrule
\end{tabular}

\end{table}

\subsection{Scenario composition and request placement}\label{app:tpos}
The $2{,}000$ scenarios comprise $1{,}500$ three-human and $500$ four-human conversations, each with one assistant.
Generation assigns T to early-middle, middle or late-middle position buckets in $667$, $667$ and $666$ scenarios, respectively.
Every scenario contains exactly one T, which is never first and is followed by further conversation.
There are $968$ scenarios with an N4 event after T.
An N4 question is addressed to Aria by design but is followed by a human answer or instruction that the assistant need not answer; it is not a second T.
The position constraint and the explicit/implicit addressing assignment are separate controls.

\subsection{Speech synthesis}\label{app:audiobody}
Qwen3-TTS synthesises human utterances separately using a fixed speaker-to-voice map.
A, B, C and D are identifiers assigned to the speakers within each scenario, not personal names.
They select the voices \texttt{ryan}, \texttt{aiden}, \texttt{eric} and \texttt{dylan}, respectively, with clips stored at $24$\,kHz.
Names mentioned in an utterance do not introduce additional speakers or voices.
Each manifest record contains the scenario and pair IDs, speaker, turn text, event label, audio path, duration and T gold answer.

\subsection{Paired edits and audio accounting}\label{app:scale}
The addressing rewrite changes only T in $1{,}958$ pairs and also changes the preceding turn in $42$ pairs.
The final scripts, including N4 edits, differ only at T in $1{,}661$ pairs; $339$ pairs have additional text differences, including $16$ with unequal turn counts.
All pairs retain the same T gold answer.

The two addressing conditions contain $2{,}000$ scenarios each.
The final human clips total $105.55$\,h across all $4{,}000$ conversations, counting clips reused between versions each time they occur.
Adding $5.37$\,h of $0.4$\,s inter-turn gaps gives $110.93$\,h of constructed conversations.
Of the clip duration, $44.45$\,h is reused between versions, leaving $61.10$\,h of distinct clips.
These durations exclude the spoken duty preamble and model-dependent waits during evaluation.

\subsection{Duration statistics}\label{app:audio}
Tables~\ref{tab:audioscene} and~\ref{tab:audioturn} report scene and utterance durations by T addressing condition; Table~\ref{tab:audioaddrturn} focuses on the T requests.
The ``All turns'' rows in Table~\ref{tab:audioturn} sum T, N1--N3, N4\textsubscript{Q} and N4\textsubscript{R}, excluding silence inserted between clips.
Totals are computed before rounding the displayed values.
Named and unnamed T requests average $8.18$\,s and $8.58$\,s, respectively.
These are descriptive duration comparisons, not a separate causal estimate of the effect of wording.

Scene-level durations in Table~\ref{tab:audioscene} include $0.4$\,s gaps between consecutive human clips.
Tables~\ref{tab:audioturn} and~\ref{tab:audioaddrturn} use only the individual clip durations, with no added inter-turn silence.

\begin{table}[htbp]
\caption{Scene duration by T addressing condition, in seconds except total hours.
p5/p95 denote the fifth/ninety-fifth percentiles.}
\label{tab:audioscene}
\centering
\small
\begin{tabular}{@{}lrrrrrrr@{}}
\toprule
Addressing & Scenarios & Total (h) & Mean & Median & p5 & p95 & Max \\
\midrule
Explicit & $2{,}000$ & $55.27$ & $99.5$ & $98.2$ & $79.4$ & $123.0$ & $297.6$ \\
Implicit & $2{,}000$ & $55.66$ & $100.2$ & $98.9$ & $79.6$ & $123.8$ & $298.0$ \\
\bottomrule
\end{tabular}

\end{table}

\begin{table}[htbp]
\caption{Turn duration by type and T addressing condition, in seconds except total minutes.
Totals exclude inter-turn silence.}
\label{tab:audioturn}
\centering
\small
\begin{tabular}{@{}llrrrrr@{}}
\toprule
Addressing & Turn type & $n$ & Total (min) & Mean & Median & Max \\
\midrule
Explicit & T (request) & $2{,}000$ & $272.6$ & $8.18$ & $7.84$ & $24.16$ \\
 & N1 & $3{,}041$ & $387.0$ & $7.64$ & $7.28$ & $188.72$ \\
 & N2 & $14{,}052$ & $1754.1$ & $7.49$ & $7.12$ & $49.60$ \\
 & N3 & $3{,}267$ & $427.5$ & $7.85$ & $7.44$ & $41.92$ \\
 & N4\textsubscript{Q} & $1{,}911$ & $128.8$ & $4.04$ & $3.68$ & $12.64$ \\
 & N4\textsubscript{R} & $1{,}911$ & $184.8$ & $5.80$ & $5.36$ & $27.60$ \\
\cmidrule(lr){2-7}
 & All turns & $26{,}182$ & $3154.9$ & $7.23$ & $6.96$ & $188.72$ \\
\addlinespace[2pt]
Implicit & T (request) & $2{,}000$ & $285.8$ & $8.58$ & $8.24$ & $24.40$ \\
 & N1 & $3{,}039$ & $387.6$ & $7.65$ & $7.28$ & $188.72$ \\
 & N2 & $14{,}054$ & $1760.1$ & $7.51$ & $7.12$ & $49.60$ \\
 & N3 & $3{,}267$ & $428.1$ & $7.86$ & $7.44$ & $41.92$ \\
 & N4\textsubscript{Q} & $1{,}911$ & $129.8$ & $4.08$ & $3.68$ & $22.96$ \\
 & N4\textsubscript{R} & $1{,}911$ & $186.8$ & $5.86$ & $5.44$ & $27.60$ \\
\cmidrule(lr){2-7}
 & All turns & $26{,}182$ & $3178.3$ & $7.28$ & $6.96$ & $188.72$ \\
\addlinespace[2pt]
\bottomrule
\end{tabular}

\end{table}

\begin{table}[htbp]
\caption{T-utterance duration by addressing condition.
Mean and median are in seconds; chars denotes mean character count.}
\label{tab:audioaddrturn}
\centering
\small
\begin{tabular}{@{}lrrrr@{}}
\toprule
Addressing & $n$ & Mean & Median & Chars \\
\midrule
Explicit & $2{,}000$ & $8.18$ & $7.84$ & $111.6$ \\
Implicit & $2{,}000$ & $8.58$ & $8.24$ & $123.3$ \\
\bottomrule
\end{tabular}

\end{table}

\section{Runtime protocol and scoring}

\subsection{Shared scheduler and duty instruction}\label{app:schedule}
Each episode starts from fresh model state with the $31.9$\,s spoken duty preamble, followed by $0.5$\,s of silence before the first human turn.
Between ordinary human turns, the scheduler plays $0.5$\,s of silence after one clip ends and before the next begins.
After T, the model has $5$\,s to start speaking; once speech is present, playback waits for $2$\,s of confirmed silence or a $60$\,s cap.
The cap starts at a new onset, or at T's end if the model is already speaking.
After N4\textsubscript{Q}, the scheduler always waits $3$\,s and then plays N4\textsubscript{R}, regardless of whether the model has finished.
The same settings apply to all eight systems.

The spoken instruction, also used verbatim by the text reference, is:
\begin{center}
\setlength{\fboxsep}{8pt}
\fbox{%
\begin{minipage}{\dimexpr\linewidth-2\fboxsep-2\fboxrule\relax}
You are the voice assistant Aria, that is A R I A, Aria.
You are in a room with several people.
Only respond when someone speaks directly to you, Aria, and asks something you can answer.
When people are talking to each other or to another device, or thinking out loud, or when the question they asked you has already been answered by someone else, stay silent and do not speak.
\end{minipage}%
}
\end{center}

\subsection{Fresh onset and response presence}\label{app:m1strict}
Table~\ref{tab:m1strict} partitions T requests into fresh onset, speech already active at request completion, and no response within the deadline.
Response presence is the sum of the first two rates.
A response exceeding the $60$\,s cap still belongs to its original onset category; duration and initiation are separate outcomes.

\begin{table}[htbp]
\caption{Fresh speech, continuation and silence across the $2{,}000$ T requests per addressing condition.}
\label{tab:m1strict}
\centering
\small
\resizebox{\textwidth}{!}{
\begin{tabular}{@{}llrrrrr@{}}
\toprule
Model & Addressing & Fresh onset & Already speaking & Miss & \shortstack{Fresh-onset\\response rate} & \shortstack{Response\\presence} \\
\midrule
MiniCPM-o 4.5 & Explicit & $1{,}896$ & $14$ & $90$ & $0.9480$ & $0.9550$ \\
 & Implicit & $1{,}880$ & $6$ & $114$ & $0.9400$ & $0.9430$ \\
\addlinespace[2pt]
Moshi & Explicit & $1{,}268$ & $362$ & $370$ & $0.6340$ & $0.8150$ \\
 & Implicit & $1{,}339$ & $305$ & $356$ & $0.6695$ & $0.8220$ \\
\addlinespace[2pt]
FLM-Audio & Explicit & $1{,}077$ & $824$ & $99$ & $0.5385$ & $0.9505$ \\
 & Implicit & $1{,}087$ & $822$ & $91$ & $0.5435$ & $0.9545$ \\
\addlinespace[2pt]
Voila & Explicit & $1{,}603$ & $7$ & $390$ & $0.8015$ & $0.8050$ \\
 & Implicit & $1{,}625$ & $9$ & $366$ & $0.8125$ & $0.8170$ \\
\addlinespace[2pt]
Freeze-Omni & Explicit & $29$ & $56$ & $1{,}915$ & $0.0145$ & $0.0425$ \\
 & Implicit & $20$ & $71$ & $1{,}909$ & $0.0100$ & $0.0455$ \\
\addlinespace[2pt]
Realtime-Venus-Audio & Explicit & $1{,}664$ & $158$ & $178$ & $0.8320$ & $0.9110$ \\
 & Implicit & $1{,}684$ & $154$ & $162$ & $0.8420$ & $0.9190$ \\
\addlinespace[2pt]
Moshika-RL-Seamless & Explicit & $1{,}755$ & $231$ & $14$ & $0.8775$ & $0.9930$ \\
 & Implicit & $1{,}816$ & $172$ & $12$ & $0.9080$ & $0.9940$ \\
\addlinespace[2pt]
PersonaPlex & Explicit & $1{,}511$ & $468$ & $21$ & $0.7555$ & $0.9895$ \\
 & Implicit & $1{,}568$ & $413$ & $19$ & $0.7840$ & $0.9905$ \\
\addlinespace[2pt]
\bottomrule
\end{tabular}
}
\end{table}

\subsection{Semantic scoring}\label{app:judgefail}
For conditional answer accuracy, Qwen3-ASR-1.7B transcribes the response and Claude Opus 5 compares it with the gold answer, ignoring filler and wording differences.
An empty response transcript is scored as incorrect.

For silence preservation, a window with no model speech detected by VAD passes directly, without ASR or an Opus call.
Only windows with detected model speech require transcription and semantic classification.
All detected model-speech fragments inside one window are joined and transcribed together.
Claude Opus 5 receives that text, up to three human turns ending at the current turn, the total model speech duration and its overlap with human speech.

For silence preservation, a brief acknowledgement that adds no information and answers no question is allowed (\texttt{BACKCHANNEL}).
Answering, explaining, advising, asking a new question, or repeating acknowledgements until they hold the floor fails this metric (\texttt{SUBSTANTIVE\_INTRUSION}).
Duration alone does not determine whether speech is allowed: even a short answer such as ``Nine dollars'' counts as an intrusion.

An empty transcript after detected speech does not establish that the model was silent or produced an allowed acknowledgement.
The window therefore fails silence preservation.
These are scoring decisions in the absence of a content classification, not confirmed substantive intrusions.

Table~\ref{tab:judgefail} counts only N1--N3 windows.
Its columns report windows where VAD detects model speech, the subset with an empty transcript, and the subset with a non-empty transcript but no valid judge label.
For example, in the explicit condition Voila has detected speech in $4{,}956$ windows; $76$ have no transcript, and all remaining windows receive a judge label.
Treating those $76$ windows as passes would add $76/20{,}360\approx0.00373$ to Voila's silence preservation, or $0.37$ percentage points, raising it from $83.97\%$ to $84.34\%$ without changing the model ordering.

\begin{table}[htbp]
\caption{Window counts at each stage of silence-preservation scoring.}
\label{tab:judgefail}
\centering
\small
\begin{tabular}{@{}llrrr@{}}
\toprule
Model & Addressing & \shortstack{VAD detects\\speech} & \shortstack{Empty ASR\\transcript} & \shortstack{No valid\\judge label} \\
\midrule
MiniCPM-o 4.5 & Explicit & $2{,}099$ & $7$ & $0$ \\
 & Implicit & $1{,}988$ & $8$ & $0$ \\
\addlinespace[2pt]
Moshi & Explicit & $18{,}626$ & $125$ & $0$ \\
 & Implicit & $18{,}654$ & $118$ & $0$ \\
\addlinespace[2pt]
FLM-Audio & Explicit & $12{,}495$ & $36$ & $0$ \\
 & Implicit & $12{,}414$ & $41$ & $0$ \\
\addlinespace[2pt]
Voila & Explicit & $4{,}956$ & $76$ & $0$ \\
 & Implicit & $4{,}893$ & $60$ & $0$ \\
\addlinespace[2pt]
Freeze-Omni & Explicit & $3{,}605$ & $67$ & $0$ \\
 & Implicit & $3{,}687$ & $72$ & $0$ \\
\addlinespace[2pt]
Realtime-Venus-Audio & Explicit & $9{,}278$ & $99$ & $0$ \\
 & Implicit & $9{,}159$ & $101$ & $0$ \\
\addlinespace[2pt]
Moshika-RL-Seamless & Explicit & $20{,}058$ & $229$ & $0$ \\
 & Implicit & $20{,}052$ & $206$ & $0$ \\
\addlinespace[2pt]
PersonaPlex & Explicit & $20{,}189$ & $895$ & $0$ \\
 & Implicit & $20{,}194$ & $853$ & $0$ \\
\addlinespace[2pt]
\bottomrule
\end{tabular}

\end{table}

\subsection{Human validation of data and semantic scoring}\label{app:humanvalidation}
Six students who understand spoken and written English evaluate the sampled materials.
Each item is assessed by two reviewers; when their judgments differ, those same two reviewers discuss the item and determine its final label.
We compare these final labels with the original Opus~5 verdicts to assess semantic scoring reliability.

For construction validation, we randomly sample $50$ scenario pairs: $38$ with three human speakers and $12$ with four, including both addressing versions of each pair.
These provide $100$ conversation items containing $1{,}282$ human-turn occurrences and $95$ N4 question--resolution pairs.
Reviewers check whether T and N4\textsubscript{Q} address Aria, whether T has a unique answer supported by the preceding dialogue, whether the supplied gold answer is supported, whether N4\textsubscript{R} answers or cancels its request, and whether N1--N3 turns require silence rather than a substantive response.
T answerability is judged using only the request and preceding dialogue, before revealing the gold answer or subsequent turns.
N4\textsubscript{Q} is checked for its addressee, without requiring an answer inferable from the preceding dialogue.
Reviewers also check whether Qwen3-TTS speech is intelligible and preserves the script's meaning.

To validate model evaluation, we separately sample outputs from the full benchmark runs of the original five systems.
Each of the five models is evaluated on $2{,}000$ scenarios per addressing condition; from its available outputs in each condition, we select ten T responses and ten N1--N3 speech outputs.
This gives $5\times2\times10=100$ outputs for each semantic scoring task, sampled independently of the $50$ conversation pairs above.
Within each model and condition, we randomly draw five T responses labelled correct and five labelled incorrect by Opus~5; for N1--N3 outputs, we draw five labelled acknowledgement and five labelled intrusion.
When a label has fewer than five examples, we include all of them and randomly fill the remaining slots from the other label.
Reviewers check whether Qwen3-ASR-1.7B transcripts preserve the meaning of these $200$ output clips and assess answer correctness or acknowledgement versus intrusion for comparison with Opus~5.
Together with the $100$ conversation items, this gives $300$ review items, each assessed by two reviewers, for $600$ initial assessments.

\begin{table}[htbp]
\caption{Human validation of sampled conversations and model outputs.}
\label{tab:humanvalidation}
\centering
\small
\begingroup
\setlength{\tabcolsep}{5pt}
\renewcommand{\arraystretch}{1.12}
\begin{tabularx}{\textwidth}{@{}>{\raggedright\arraybackslash}Xrr@{}}
\toprule
\textbf{Check} & \textbf{Sample size} & \textbf{Rate} \\
\midrule
\multicolumn{3}{@{}l}{\textbf{Construction: Opus~5 and Qwen3-TTS}} \\
T addresses Aria: explicit & $50$ & $100\%$ \\
T addresses Aria: implicit & $50$ & $100\%$ \\
T has a unique answer supported by preceding dialogue & $100$ & $100\%$ \\
Gold answer is correct and supported & $100$ & $98\%$ \\
N4\textsubscript{Q} addresses Aria & $95$ & $100\%$ \\
N4\textsubscript{R} answers or cancels its request & $95$ & $100\%$ \\
N1--N3 turn requires no substantive response & $992$ & $99.70\%$ \\
TTS is intelligible and preserves script meaning & $1{,}282$ & $100\%$ \\
\midrule
\multicolumn{3}{@{}l}{\textbf{Evaluation: Qwen3-ASR-1.7B and Opus~5}} \\
ASR preserves model output meaning & $200$ & $100\%$ \\
Agreement with Opus~5: T answer correctness & $100$ & $99\%$ \\
Agreement with Opus~5: acknowledgement versus intrusion & $100$ & $98\%$ \\
\bottomrule
\end{tabularx}
\endgroup

\end{table}

Table~\ref{tab:humanvalidation} summarises validation of benchmark construction and model evaluation.
Construction checks achieve $98\%$--$100\%$ pass rates, including intelligible and meaning-preserving TTS for all checked turns.
For evaluation, all checked ASR transcripts preserve meaning, and reviewers' final labels agree with Opus~5 on $99\%$ of answer-correctness judgments and $98\%$ of acknowledgement-versus-intrusion judgments.
These results support the reliability of both stages on the sampled items.

\subsection{Request timing and response quality}\label{app:requesttime}
Table~\ref{tab:requesttime} gives the complete results for the analysis in Sec.~\ref{sec:requesttime}.
Elapsed time $t$ runs from the first human turn to T onset, including intervening speech and scheduled gaps but excluding the duty preamble; the observed range is $14.50$--$263.97$\,s.
We sort the $4{,}000$ requests by this time and divide them into three groups with approximately equal sample sizes.
The one-third and two-thirds quantiles give boundaries of $40.224$ and $54.656$\,s, placing $1{,}330$, $1{,}335$ and $1{,}335$ requests in the earlier, middle and later groups.
All models use the same grouping, determined from input timing rather than model scores.
Conditional answer accuracy uses response-present requests as its denominator; the two response rates use all requests in the group.

\begin{table}[p]
\caption{Performance by request time; all rates are percentages.}
\label{tab:requesttime}
\centering
\footnotesize
\begingroup
\setlength{\tabcolsep}{3pt}
\renewcommand{\arraystretch}{1.06}
\begin{tabularx}{\textwidth}{@{}l l l r *{3}{>{\raggedleft\arraybackslash}X}@{}}
\toprule
Model & Addressing & T time & Requests & \shortstack{Fresh\\onset} & \shortstack{Response\\presence} & \shortstack{Answer\\accuracy} \\
\midrule
MiniCPM-o 4.5 & Explicit & Earlier & 670 & 96.12 & 97.16 & 48.85 \\
 &  & Middle & 665 & 93.68 & 94.59 & 43.72 \\
 &  & Later & 665 & 94.59 & 94.74 & 43.17 \\
 & Implicit & Earlier & 660 & 95.15 & 95.61 & 51.19 \\
 &  & Middle & 670 & 92.99 & 93.28 & 44.00 \\
 &  & Later & 670 & 93.88 & 94.03 & 44.13 \\
\addlinespace[2pt]
Moshi & Explicit & Earlier & 670 & 62.39 & 81.04 & 1.66 \\
 &  & Middle & 665 & 64.06 & 80.30 & 1.12 \\
 &  & Later & 665 & 63.76 & 83.16 & 1.45 \\
 & Implicit & Earlier & 660 & 66.82 & 80.76 & 1.69 \\
 &  & Middle & 670 & 68.96 & 84.18 & 1.42 \\
 &  & Later & 670 & 65.07 & 81.64 & 1.65 \\
\addlinespace[2pt]
FLM-Audio & Explicit & Earlier & 670 & 67.61 & 94.03 & 0.16 \\
 &  & Middle & 665 & 46.02 & 96.39 & 0.00 \\
 &  & Later & 665 & 47.82 & 94.74 & 0.16 \\
 & Implicit & Earlier & 660 & 66.97 & 95.30 & 0.00 \\
 &  & Middle & 670 & 47.16 & 95.67 & 0.16 \\
 &  & Later & 670 & 49.10 & 95.37 & 0.00 \\
\addlinespace[2pt]
Voila & Explicit & Earlier & 670 & 84.93 & 85.52 & 2.27 \\
 &  & Middle & 665 & 79.10 & 79.55 & 0.76 \\
 &  & Later & 665 & 76.39 & 76.39 & 0.59 \\
 & Implicit & Earlier & 660 & 85.45 & 86.21 & 2.28 \\
 &  & Middle & 670 & 81.04 & 81.34 & 1.65 \\
 &  & Later & 670 & 77.31 & 77.61 & 0.96 \\
\addlinespace[2pt]
Freeze-Omni & Explicit & Earlier & 670 & 0.15 & 1.04 & 0.00 \\
 &  & Middle & 665 & 2.71 & 5.56 & 0.00 \\
 &  & Later & 665 & 1.50 & 6.17 & 0.00 \\
 & Implicit & Earlier & 660 & 0.15 & 0.91 & 0.00 \\
 &  & Middle & 670 & 1.34 & 5.97 & 0.00 \\
 &  & Later & 670 & 1.49 & 6.72 & 0.00 \\
\addlinespace[2pt]
Realtime-Venus-Audio & Explicit & Earlier & 670 & 82.99 & 91.19 & 27.17 \\
 &  & Middle & 665 & 83.91 & 91.73 & 20.82 \\
 &  & Later & 665 & 82.71 & 90.38 & 22.30 \\
 & Implicit & Earlier & 660 & 85.91 & 91.82 & 28.38 \\
 &  & Middle & 670 & 82.84 & 91.94 & 19.97 \\
 &  & Later & 670 & 83.88 & 91.94 & 21.27 \\
\addlinespace[2pt]
Moshika-RL-Seamless & Explicit & Earlier & 670 & 87.61 & 99.55 & 1.95 \\
 &  & Middle & 665 & 88.87 & 98.95 & 1.98 \\
 &  & Later & 665 & 86.77 & 99.40 & 1.97 \\
 & Implicit & Earlier & 660 & 90.45 & 99.55 & 2.74 \\
 &  & Middle & 670 & 90.45 & 99.40 & 2.25 \\
 &  & Later & 670 & 91.49 & 99.25 & 1.65 \\
\addlinespace[2pt]
PersonaPlex & Explicit & Earlier & 670 & 74.63 & 99.10 & 3.01 \\
 &  & Middle & 665 & 77.89 & 99.25 & 1.52 \\
 &  & Later & 665 & 74.14 & 98.50 & 1.98 \\
 & Implicit & Earlier & 660 & 75.61 & 98.79 & 1.38 \\
 &  & Middle & 670 & 81.34 & 99.70 & 2.40 \\
 &  & Later & 670 & 78.21 & 98.66 & 1.36 \\
\bottomrule
\end{tabularx}
\endgroup

\end{table}

\section{N4 stopping analysis}

\subsection{Eligibility and stopping rule}\label{app:stopcrit}
Window response covers any detected model speech from N4\textsubscript{Q} onset through its $3$\,s answering window.
Table~\ref{tab:ayfunnel} separates three cases: the model is already speaking when N4\textsubscript{Q} begins, starts speaking during N4\textsubscript{Q}, or first speaks in the $3$\,s interval after N4\textsubscript{Q} ends.
Only the third group can enter answering-window yield, and only if the model is still speaking when N4\textsubscript{R} begins.
Models that finish speaking before N4\textsubscript{R} begins are excluded from the stopping denominator.

Let $t_R$ be the time when the N4\textsubscript{R} human audio clip starts playing, and $t_E$ the time when that clip ends plus $0.5$\,s of inter-turn silence.
An eligible event passes if the model is silent at $t_R+2$\,s.
If $t_E$ is later than this deadline, the model must also remain silent from the deadline until $t_E$.
The deadline stays at $t_R+2$\,s even when the human clip and its following silence end earlier.
For example, if N4\textsubscript{R} lasts $1$\,s, then $t_E=t_R+1.5$\,s; an eligible model that stops at $t_R+1.8$\,s and is silent at the deadline passes, although it stops after the human clip ends.
A model still speaking at $t_R+2$\,s fails.
Stopping is computed from the detector's speech segments, which merge gaps shorter than $800$\,ms.
Answer content is not part of this stopping test.

PersonaPlex is already speaking when N4\textsubscript{Q} begins in $1{,}599$ explicit-condition events and $1{,}606$ implicit-condition events.
It starts during N4\textsubscript{Q} in another $124$ and $121$ events, respectively, leaving $76$ and $66$ eligible events.
Conditional stopping rates must therefore be read together with the counts in Table~\ref{tab:ayfunnel}.

\begin{table}[t]
\caption{Events entering answering-window yield.
Q and R denote the N4 question and resolution; the answering interval is the $3$\,s gap between them.}
\label{tab:ayfunnel}
\centering
\footnotesize
\setlength{\tabcolsep}{1.8pt}
\resizebox{\textwidth}{!}{
\begin{tabular}{@{}llrrrrrr@{}}
\toprule
& & & \multicolumn{3}{c}{Model speech relative to Q} & \multicolumn{2}{c}{Among starts after Q} \\
\cmidrule(lr){4-6}\cmidrule(lr){7-8}
Model & Addressing & \shortstack{Window\\response hits} & \shortstack{Already speaking\\at Q start} & \shortstack{Starts\\during Q} & \shortstack{Starts after Q\\within 3\,s} & \shortstack{Silent\\at R start} & \shortstack{Still speaking\\at R start} \\
\midrule
MiniCPM-o 4.5 & Explicit & $1{,}818$ & $6$ & $61$ & $1{,}751$ & $90$ & $1{,}661$ \\
 & Implicit & $1{,}815$ & $2$ & $61$ & $1{,}752$ & $96$ & $1{,}656$ \\
\addlinespace[2pt]
Moshi & Explicit & $1{,}764$ & $914$ & $421$ & $429$ & $208$ & $221$ \\
 & Implicit & $1{,}779$ & $905$ & $427$ & $447$ & $206$ & $241$ \\
\addlinespace[2pt]
FLM-Audio & Explicit & $1{,}897$ & $617$ & $284$ & $996$ & $19$ & $977$ \\
 & Implicit & $1{,}900$ & $598$ & $268$ & $1{,}034$ & $16$ & $1{,}018$ \\
\addlinespace[2pt]
Voila & Explicit & $1{,}495$ & $16$ & $254$ & $1{,}225$ & $196$ & $1{,}029$ \\
 & Implicit & $1{,}500$ & $15$ & $261$ & $1{,}224$ & $192$ & $1{,}032$ \\
\addlinespace[2pt]
Freeze-Omni & Explicit & $345$ & $206$ & $44$ & $95$ & $2$ & $93$ \\
 & Implicit & $343$ & $205$ & $48$ & $90$ & $3$ & $87$ \\
\addlinespace[2pt]
Realtime-Venus-Audio & Explicit & $1{,}764$ & $145$ & $658$ & $961$ & $41$ & $920$ \\
 & Implicit & $1{,}779$ & $156$ & $632$ & $991$ & $37$ & $954$ \\
\addlinespace[2pt]
Moshika-RL-Seamless & Explicit & $1{,}911$ & $910$ & $463$ & $538$ & $322$ & $216$ \\
 & Implicit & $1{,}911$ & $879$ & $452$ & $580$ & $338$ & $242$ \\
\addlinespace[2pt]
PersonaPlex & Explicit & $1{,}911$ & $1{,}599$ & $124$ & $188$ & $112$ & $76$ \\
 & Implicit & $1{,}911$ & $1{,}606$ & $121$ & $184$ & $118$ & $66$ \\
\addlinespace[2pt]
\bottomrule
\end{tabular}
}
\end{table}

\subsection{Stopping outcomes}\label{app:ay}
Table~\ref{tab:ayfailure} accounts for every scored event: successful stopping, speech spanning the deadline, or silence at the deadline followed by later speech.
In the explicit condition, FLM-Audio is speaking at the deadline in $968$ of $977$ events.
Moshi is silent at the deadline but speaks again before observation ends in $98$ of its $179$ failures.
These events fail the subsequent-silence check even though the model briefly paused.

\begin{table}[htbp]
\caption{Stopping outcomes among scored N4 events.
\emph{Still voicing}: speaking at the $2$\,s deadline; \emph{resumed}: silent then, but speaking again before observation ends.}
\label{tab:ayfailure}
\centering
\small
\begin{tabular}{@{}llrrrr@{}}
\toprule
Model & Addressing & Scored & Yielded & Still voicing & Resumed \\
\midrule
MiniCPM-o 4.5 & Explicit & $1{,}661$ & $935$ & $722$ & $4$ \\
 & Implicit & $1{,}656$ & $925$ & $727$ & $4$ \\
\addlinespace[2pt]
Moshi & Explicit & $221$ & $42$ & $81$ & $98$ \\
 & Implicit & $241$ & $51$ & $92$ & $98$ \\
\addlinespace[2pt]
FLM-Audio & Explicit & $977$ & $8$ & $968$ & $1$ \\
 & Implicit & $1{,}018$ & $6$ & $1{,}012$ & $0$ \\
\addlinespace[2pt]
Voila & Explicit & $1{,}029$ & $934$ & $43$ & $52$ \\
 & Implicit & $1{,}032$ & $926$ & $40$ & $66$ \\
\addlinespace[2pt]
Freeze-Omni & Explicit & $93$ & $6$ & $83$ & $4$ \\
 & Implicit & $87$ & $2$ & $77$ & $8$ \\
\addlinespace[2pt]
Realtime-Venus-Audio & Explicit & $920$ & $334$ & $571$ & $15$ \\
 & Implicit & $954$ & $351$ & $591$ & $12$ \\
\addlinespace[2pt]
Moshika-RL-Seamless & Explicit & $216$ & $12$ & $50$ & $154$ \\
 & Implicit & $242$ & $11$ & $82$ & $149$ \\
\addlinespace[2pt]
PersonaPlex & Explicit & $76$ & $1$ & $14$ & $61$ \\
 & Implicit & $66$ & $2$ & $16$ & $48$ \\
\addlinespace[2pt]
\bottomrule
\end{tabular}

\end{table}

\section{Addressing and transcript-reference controls}

\subsection{Paired addressing comparison}\label{app:nowake}
Each of the $2{,}000$ scenarios is evaluated in both an explicit and an implicit addressing version, giving two observations of the same model on the corresponding T request.
Here, a response means detected model speech in T's response window, including speech already underway; answer correctness is evaluated separately.
Table~\ref{tab:pairedaddressing} places each pair in one of four groups: speech in both versions, only in the explicit version, only in the implicit version, or in neither.
These four counts sum to $2{,}000$ for each model.

The explicit-minus-implicit difference measures the net change in how often the model speaks between the paired versions.
It is the explicit-only count minus the implicit-only count, divided by $2{,}000$; the table multiplies this value by $100$ to express it in percentage points.
For MiniCPM-o, $98$ pairs favour the explicit version and $74$ favour the implicit version, giving $24$ more responses overall, or $1.20$ percentage points.
This net difference can be small even when the model changes its response on many individual pairs.

The exact two-sided McNemar test evaluates whether the two directions of change are equally likely.
The significance level $\alpha=0.05$ is the threshold applied to the table's $p$ column: only $p<0.05$ is called statistically significant.
MiniCPM-o's $p=0.0792$ exceeds this threshold, so its $98$ versus $74$ split does not provide sufficient evidence of unequal response probabilities.
The same holds for all eight models; insufficient evidence of a difference does not establish that the conditions are equivalent.

\begin{table}[htbp]
\caption{Paired addressing comparisons over $2{,}000$ scenario pairs.}
\label{tab:pairedaddressing}
\centering
\footnotesize
\setlength{\tabcolsep}{4pt}
\begin{tabular}{@{}lrrrrrr@{}}
\toprule
& \multicolumn{4}{c}{Model speech after T in each pair} & & \\
\cmidrule(lr){2-5}
Model & \shortstack{Both\\versions} & \shortstack{Explicit\\only} & \shortstack{Implicit\\only} & Neither & \shortstack{Explicit $-$ implicit\\(percentage points)} & $p$ \\
\midrule
MiniCPM-o 4.5 & $1{,}812$ & $98$ & $74$ & $16$ & $+1.20$ & $0.0792$ \\
Moshi & $1{,}354$ & $276$ & $290$ & $80$ & $-0.70$ & $0.5848$ \\
FLM-Audio & $1{,}816$ & $85$ & $93$ & $6$ & $-0.40$ & $0.5999$ \\
Voila & $1{,}333$ & $277$ & $301$ & $89$ & $-1.20$ & $0.3387$ \\
Freeze-Omni & $29$ & $56$ & $62$ & $1{,}853$ & $-0.30$ & $0.6455$ \\
Realtime-Venus-Audio & $1{,}674$ & $148$ & $164$ & $14$ & $-0.80$ & $0.3958$ \\
Moshika-RL-Seamless & $1{,}974$ & $12$ & $14$ & $0$ & $-0.10$ & $0.8450$ \\
PersonaPlex & $1{,}960$ & $19$ & $21$ & $0$ & $-0.10$ & $0.8746$ \\
\bottomrule
\end{tabular}

\end{table}

\subsection{Text-based evaluation of response decisions and answers}\label{app:textmain}
We evaluate whether Gemini~3.1~Pro chooses to respond to the right utterances and answers T requests correctly when given text instead of audio.
For each evaluated utterance, it receives that utterance, the duty instruction in App.~\ref{app:schedule} and preceding speaker-attributed dialogue, then chooses \texttt{RESPOND} or \texttt{SILENT}.
Decisions are queried independently using the script history, without future utterances or previous model-generated answers.

The first four result columns in Table~\ref{tab:textmain} correspond to the following tasks:
\begin{itemize}
\item \textbf{T requests}: response presence is the fraction receiving a \texttt{RESPOND} decision.
For these requests only, Gemini generates an answer, and a separate Gemini call compares it with the gold answer to compute conditional answer accuracy.
\item \textbf{N1, N2 and N3 utterances}: silence preservation is the fraction receiving a \texttt{SILENT} decision.
\item \textbf{N4\textsubscript{Q} questions}: window response is the fraction receiving a \texttt{RESPOND} decision.
No answer is generated for these questions, and this text decision does not measure whether speech begins within three seconds.
\end{itemize}
The text evaluation produces no speech timeline, so it does not measure fresh-onset response rate or stopping after N4\textsubscript{R} begins.

The final column is a separate diagnostic: Gemini is asked whether each N4\textsubscript{Q} question addresses Aria, regardless of whether it can answer.
All these questions are intended for Aria; the column reports how often Gemini identifies that addressee, and does not enter any of the four preceding metrics.

The explicit and implicit rows group scenarios by T's addressing form; their N4 and silence-preservation columns do not imply that every turn in those scenarios has that addressing form.
Across all $2{,}000$ matched pairs, explicit and implicit response rates are $0.9470$ and $0.3040$, giving a difference of $0.6430$.

To examine consistency across runs, we compare two text evaluations of Gemini~3.1~Pro on the same scenario versions with identical duty instructions.
The runs agree on $97.2\%$ of the $48{,}542$ decisions to respond or remain silent.
For Gemini, response presence, silence preservation and window response differ by at most $0.95$ percentage points between runs.
Its conditional answer accuracy changes from $90.39\%$ to $97.02\%$ on explicit requests, a $6.63$-percentage-point increase.
On implicit requests it changes from $79.11\%$ to $81.03\%$, a $1.92$-percentage-point increase.
The paired addressing differences are $0.6430$ and $0.6455$, so the large addressing contrast is reproduced despite the answer-accuracy variation.

\begin{table}[tb]
\caption{Transcript-conditioned results for Gemini~3.1~Pro.
Bold marks the higher value across conditions in columns marked $\uparrow$.}
\label{tab:textmain}
\centering
\small
\setlength{\tabcolsep}{4pt}
\begin{tabular}{@{}lccccc@{}}
\toprule
Addressing & \shortstack{Response\\presence (T)} & \shortstack{Conditional\\answer accuracy}$\uparrow$ & \shortstack{Silence\\preservation}$\uparrow$ & \shortstack{Window\\response (N4\textsubscript{Q})} & \shortstack{N4\textsubscript{Q}: Gemini\\identifies Aria} \\
\midrule
Explicit & $0.9470$ & $\mathbf{0.9039}$ & $0.9537$ & $0.7425$ & $0.9236$ \\
Implicit & $0.3040$ & $0.7911$ & $\mathbf{0.9934}$ & $0.7609$ & $0.9299$ \\
\bottomrule
\end{tabular}

\end{table}

\subsection{Silence preservation by turn type}\label{app:textsp}
Table~\ref{tab:textsp} breaks down text-reference silence preservation by taxonomy label.

\begin{table}[htbp]
\caption{Transcript-conditioned silence preservation by turn type.}
\label{tab:textsp}
\centering
\small
\begin{tabular}{@{}lcccc@{}}
\toprule
Addressing & N1 & N2 & N3 & All \\
\midrule
Explicit & $0.9231$ & $0.9715$ & $0.9057$ & $0.9537$ \\
Implicit & $0.9865$ & $0.9964$ & $0.9865$ & $0.9934$ \\
\bottomrule
\end{tabular}

\end{table}

Table~\ref{tab:spact} groups the N1, N2 and N3 windows for all eight speech systems by utterance type and addressee.
Requests and questions are split into device-directed and other requests; the latter are listed as requests to a human.
Parentheses in the column headings identify the N categories represented in each group; a column need not cover a whole category.
For example, a question to another human that mentions Aria belongs to N1, so the human-directed request column contains both N1 and N2.
Statements span N1, N2 and N3, while self-talk is a subset of N3.
Failures include both substantive intrusions and empty-transcript cases, as in the main silence-preservation score.
For MiniCPM-o under explicit addressing, the device-request column contains $121$ failures among $2{,}570$ windows and the human-request column contains $65$ among $2{,}134$.
Combining their counts gives $(121+65)/(2{,}570+2{,}134)\approx3.95\%$, reported as $4.0\%$ in the main text.

\begin{table}[htbp]
\caption{Speech-system silence-preservation failure rates by utterance type and addressee.
Lower is better; bold marks the lowest value per addressing condition and column.}
\label{tab:spact}
\centering
\small
\begin{tabular}{@{}llrrrr@{}}
\toprule
& & \multicolumn{2}{c}{Carries a request\dots} & & \\
\cmidrule(lr){3-4}
Model & Addressing & \shortstack{to a device\\(N2)} & \shortstack{to a human\\(N1/N2)} & \shortstack{statement\\(N1/N2/N3)} & \shortstack{self-talk\\(N3)} \\
\midrule
MiniCPM-o 4.5 & Explicit & $\mathbf{4.7\%}$ & $\mathbf{3.0\%}$ & $\mathbf{9.2\%}$ & $\mathbf{4.9\%}$ \\
 & Implicit & $\mathbf{5.0\%}$ & $\mathbf{2.0\%}$ & $\mathbf{9.0\%}$ & $\mathbf{5.3\%}$ \\
Moshi & Explicit & $73.1\%$ & $79.3\%$ & $69.9\%$ & $77.4\%$ \\
 & Implicit & $74.9\%$ & $79.5\%$ & $70.8\%$ & $75.4\%$ \\
FLM-Audio & Explicit & $49.7\%$ & $56.8\%$ & $63.4\%$ & $59.4\%$ \\
 & Implicit & $50.1\%$ & $56.5\%$ & $63.0\%$ & $59.0\%$ \\
Voila & Explicit & $12.8\%$ & $10.0\%$ & $16.4\%$ & $21.2\%$ \\
 & Implicit & $11.7\%$ & $10.5\%$ & $15.7\%$ & $22.5\%$ \\
Freeze-Omni & Explicit & $19.3\%$ & $28.4\%$ & $13.1\%$ & $26.0\%$ \\
 & Implicit & $19.9\%$ & $29.1\%$ & $13.3\%$ & $26.4\%$ \\
Realtime-Venus-Audio & Explicit & $39.5\%$ & $40.9\%$ & $39.1\%$ & $41.2\%$ \\
 & Implicit & $38.6\%$ & $42.3\%$ & $38.4\%$ & $40.4\%$ \\
Moshika-RL-Seamless & Explicit & $54.3\%$ & $61.0\%$ & $48.1\%$ & $62.1\%$ \\
 & Implicit & $53.9\%$ & $60.9\%$ & $47.3\%$ & $62.6\%$ \\
PersonaPlex & Explicit & $95.1\%$ & $96.1\%$ & $95.8\%$ & $96.9\%$ \\
 & Implicit & $94.7\%$ & $95.9\%$ & $96.2\%$ & $96.5\%$ \\
Window count & Explicit & $2{,}570$ & $2{,}134$ & $12{,}437$ & $3{,}219$ \\
Window count & Implicit & $2{,}570$ & $2{,}134$ & $12{,}437$ & $3{,}219$ \\
\bottomrule
\end{tabular}

\end{table}

\section{Qualitative examples}

\subsection{Allowed acknowledgements and penalised speech}\label{app:cases}
Table~\ref{tab:casessilence} shows selected model outputs and the judge's reasons for allowing or penalising them in silence-requiring windows.

\begin{table}[htbp]
\caption{Examples of exempted acknowledgements and penalised speech in silence-requiring windows.}
\label{tab:casessilence}
\centering
\small
\setlength{\tabcolsep}{4pt}
\resizebox{\textwidth}{!}{
\begin{tabular}{@{}llp{0.30\textwidth}p{0.30\textwidth}@{}}
\toprule
Model & Label (duration) & What the assistant said & Judge's reason \\
\midrule
MiniCPM-o & exempt (0.61\,s) & ``Got it.'' & Brief phatic acknowledgement; answers nothing, adds no information. \\
Realtime-Venus-Audio & penalised (1.57\,s) & ``This math is wrong.'' & Asserts an evaluative claim about the math; adds content, not phatic acknowledgement. \\
Moshika-RL-Seamless & penalised (3.62\,s) & ``From campus to Ninth and Pine takes about fifteen minutes.'' & Answered a question addressed to another assistant, supplying new information over speech. \\
PersonaPlex & penalised (1.92\,s) & ``It helps with math tests, right?'' & Asks its own question, overlapping speech; not a phatic acknowledgement. \\
Moshi & exempt (0.70\,s) & ``Yeah. Yeah.'' & brief repeated acknowledgement, no added information \\
FLM-Audio & penalised (0.48\,s) & ``Ask me.'' & It prompts interaction and tries to steer conversation. \\
Voila & exempt (0.86\,s) & ``Thank you.'' & Brief phatic 'Thank you' answers nothing and adds no information. \\
Freeze-Omni & exempt (0.90\,s) & ``I'm glad to hear.'' & Brief phatic reaction to B's relief; adds no information or answer. \\
\bottomrule
\end{tabular}
}
\end{table}

\section{Statistical comparisons}

\subsection{Confidence intervals and paired tests}\label{app:stats}
These analyses compare models on the same evaluation items to determine which score differences have statistical support.
For silence preservation, all model pairs differ significantly under both addressing conditions, supporting the ordering MiniCPM-o, Voila, Freeze-Omni, Realtime-Venus-Audio, Moshika-RL-Seamless, FLM-Audio, Moshi, then PersonaPlex from highest to lowest.
The scenario-level bootstrap supports the same ordering: every $95\%$ interval for a model-pair difference lies entirely above or below zero.
For the closest pair under explicit addressing, Voila and Freeze-Omni, the interval places Voila ahead by $0.42$ to $2.53$ percentage points.
For each resampled set of scenarios, we subtract Freeze-Omni's silence-preservation rate from Voila's rate and multiply by $100$.
The 2.5th and 97.5th percentiles of these $10{,}000$ differences are $0.4221$ and $2.5326$ percentage points, rounded to the reported bounds.

For conditional answer accuracy, MiniCPM-o exceeds all other models, Realtime-Venus-Audio exceeds the remaining six, and Moshi, Voila, Moshika-RL-Seamless and PersonaPlex each exceed FLM-Audio under both addressing conditions.
The remaining comparisons are not statistically significant, so the tests do not establish a complete ranking for this metric.
In particular, PersonaPlex's higher point estimate than Moshi does not establish higher answer accuracy on their shared requests.

Table~\ref{tab:pairwise} reports exact two-sided McNemar tests; corrected $p<0.05$ indicates a significant difference.
Bonferroni correction covers $56$ tests per addressing condition: $28$ model pairs for each of the two metrics.
Silence comparisons use all common windows; answer-accuracy comparisons use only T requests where both models produced speech, so their rates can differ from the model-specific rates in Table~\ref{tab:main}.
For the silence-preservation intervals, we sample $2{,}000$ scenarios with replacement per addressing condition, keeping each scenario's windows together, and repeat this $10{,}000$ times with seed $20261026$.
The interval bounds are the 2.5th and 97.5th percentiles of the resulting score differences.

In a row labelled ``A vs B'', $r_1$ is model A's score and $r_2$ is model B's score on the same compared items; higher is better for both metrics.
``disc.'' counts items where one model passes and the other fails: one preserves silence and the other does not, or one answers correctly and the other incorrectly.
``shared'' counts T requests on which both models produced speech; it is the denominator for both answer-accuracy scores in that row.
The corrected value $p_{\mathrm{corr}}$ is the raw McNemar $p$ value multiplied by $56$, capped at $1$; values below $0.05$ indicate a statistically significant difference, while larger values do not establish equal performance.
For example, under explicit addressing FLM-Audio and Freeze-Omni both speak on $81$ T requests and neither answers any of them correctly, so they differ in correctness on no request.
Their corrected $p=1.000$ does not support an answer-accuracy difference.

\begin{table}[p]
\caption{Pairwise comparisons with Bonferroni-corrected McNemar tests.}
\label{tab:pairwise}
\centering
\scriptsize
\setlength{\tabcolsep}{3pt}
\resizebox{\textwidth}{!}{
\begin{tabular}{@{}lrrrrrrrrr@{}}
\toprule
& \multicolumn{4}{c}{Silence preservation (all silence windows)} & \multicolumn{5}{c}{Conditional answer accuracy (shared items)} \\
\cmidrule(lr){2-5}\cmidrule(lr){6-10}
Pair & r$_1$ & r$_2$ & disc. & $p_{\mathrm{corr}}$ & r$_1$ & r$_2$ & shared & disc. & $p_{\mathrm{corr}}$ \\
\midrule
\multicolumn{10}{l}{\textbf{Explicit}} \\
FLM-Audio vs Freeze-Omni & $0.3963$ & $0.8249$ & $12{,}019$ & $<10^{-300}$ & $0.0000$ & $0.0000$ & $81$ & $0$ & $1.000$ \\
FLM-Audio vs MiniCPM-o 4.5 & $0.3963$ & $0.9267$ & $11{,}532$ & $<10^{-300}$ & $0.0011$ & $0.4523$ & $1{,}813$ & $818$ & $6.4\!\times\!10^{-245}$ \\
FLM-Audio vs Moshi & $0.3963$ & $0.2753$ & $8{,}960$ & $3.5\!\times\!10^{-149}$ & $0.0000$ & $0.0136$ & $1{,}544$ & $21$ & $5.3\!\times\!10^{-5}$ \\
FLM-Audio vs Moshika-RL-Seamless & $0.3963$ & $0.4753$ & $9{,}897$ & $2.9\!\times\!10^{-57}$ & $0.0011$ & $0.0207$ & $1{,}887$ & $41$ & $4.4\!\times\!10^{-8}$ \\
FLM-Audio vs PersonaPlex & $0.3963$ & $0.0413$ & $7{,}968$ & $<10^{-300}$ & $0.0005$ & $0.0229$ & $1{,}880$ & $44$ & $2.9\!\times\!10^{-10}$ \\
FLM-Audio vs Realtime-Venus-Audio & $0.3963$ & $0.6032$ & $9{,}851$ & $<10^{-300}$ & $0.0012$ & $0.2347$ & $1{,}734$ & $407$ & $1.4\!\times\!10^{-118}$ \\
FLM-Audio vs Voila & $0.3963$ & $0.8397$ & $10{,}737$ & $<10^{-300}$ & $0.0013$ & $0.0124$ & $1{,}528$ & $21$ & $0.012$ \\
Freeze-Omni vs MiniCPM-o 4.5 & $0.8249$ & $0.9267$ & $4{,}951$ & $9.2\!\times\!10^{-195}$ & $0.0000$ & $0.5000$ & $80$ & $40$ & $1.0\!\times\!10^{-10}$ \\
Freeze-Omni vs Moshi & $0.8249$ & $0.2753$ & $13{,}131$ & $<10^{-300}$ & $0.0000$ & $0.0132$ & $76$ & $1$ & $1.000$ \\
Freeze-Omni vs Moshika-RL-Seamless & $0.8249$ & $0.4753$ & $10{,}260$ & $<10^{-300}$ & $0.0000$ & $0.0238$ & $84$ & $2$ & $1.000$ \\
Freeze-Omni vs PersonaPlex & $0.8249$ & $0.0413$ & $16{,}329$ & $<10^{-300}$ & $0.0000$ & $0.0118$ & $85$ & $1$ & $1.000$ \\
Freeze-Omni vs Realtime-Venus-Audio & $0.8249$ & $0.6032$ & $8{,}826$ & $<10^{-300}$ & $0.0000$ & $0.2024$ & $84$ & $17$ & $8.5\!\times\!10^{-4}$ \\
Freeze-Omni vs Voila & $0.8249$ & $0.8397$ & $6{,}062$ & $0.006$ & $0.0000$ & $0.0159$ & $63$ & $1$ & $1.000$ \\
MiniCPM-o 4.5 vs Moshi & $0.9267$ & $0.2753$ & $13{,}970$ & $<10^{-300}$ & $0.4522$ & $0.0141$ & $1{,}559$ & $689$ & $2.4\!\times\!10^{-198}$ \\
MiniCPM-o 4.5 vs Moshika-RL-Seamless & $0.9267$ & $0.4753$ & $10{,}601$ & $<10^{-300}$ & $0.4539$ & $0.0184$ & $1{,}899$ & $843$ & $1.2\!\times\!10^{-233}$ \\
MiniCPM-o 4.5 vs PersonaPlex & $0.9267$ & $0.0413$ & $18{,}094$ & $<10^{-300}$ & $0.4529$ & $0.0217$ & $1{,}890$ & $833$ & $1.0\!\times\!10^{-228}$ \\
MiniCPM-o 4.5 vs Realtime-Venus-Audio & $0.9267$ & $0.6032$ & $7{,}991$ & $<10^{-300}$ & $0.4494$ & $0.2388$ & $1{,}738$ & $616$ & $1.9\!\times\!10^{-50}$ \\
MiniCPM-o 4.5 vs Voila & $0.9267$ & $0.8397$ & $3{,}927$ & $3.6\!\times\!10^{-180}$ & $0.4507$ & $0.0130$ & $1{,}542$ & $695$ & $4.7\!\times\!10^{-186}$ \\
Moshi vs Moshika-RL-Seamless & $0.2753$ & $0.4753$ & $8{,}819$ & $<10^{-300}$ & $0.0142$ & $0.0198$ & $1{,}619$ & $47$ & $1.000$ \\
Moshi vs PersonaPlex & $0.2753$ & $0.0413$ & $5{,}774$ & $<10^{-300}$ & $0.0142$ & $0.0229$ & $1{,}615$ & $52$ & $1.000$ \\
Moshi vs Realtime-Venus-Audio & $0.2753$ & $0.6032$ & $10{,}725$ & $<10^{-300}$ & $0.0141$ & $0.2401$ & $1{,}487$ & $362$ & $2.9\!\times\!10^{-84}$ \\
Moshi vs Voila & $0.2753$ & $0.8397$ & $13{,}163$ & $<10^{-300}$ & $0.0159$ & $0.0106$ & $1{,}320$ & $33$ & $1.000$ \\
Moshika-RL-Seamless vs PersonaPlex & $0.4753$ & $0.0413$ & $9{,}577$ & $<10^{-300}$ & $0.0198$ & $0.0219$ & $1{,}965$ & $72$ & $1.000$ \\
Moshika-RL-Seamless vs Realtime-Venus-Audio & $0.4753$ & $0.6032$ & $9{,}768$ & $5.9\!\times\!10^{-153}$ & $0.0204$ & $0.2348$ & $1{,}810$ & $432$ & $5.3\!\times\!10^{-92}$ \\
Moshika-RL-Seamless vs Voila & $0.4753$ & $0.8397$ & $10{,}346$ & $<10^{-300}$ & $0.0200$ & $0.0125$ & $1{,}601$ & $52$ & $1.000$ \\
PersonaPlex vs Realtime-Venus-Audio & $0.0413$ & $0.6032$ & $11{,}935$ & $<10^{-300}$ & $0.0200$ & $0.2336$ & $1{,}802$ & $425$ & $1.3\!\times\!10^{-92}$ \\
PersonaPlex vs Voila & $0.0413$ & $0.8397$ & $16{,}451$ & $<10^{-300}$ & $0.0188$ & $0.0119$ & $1{,}592$ & $47$ & $1.000$ \\
Realtime-Venus-Audio vs Voila & $0.6032$ & $0.8397$ & $8{,}632$ & $<10^{-300}$ & $0.2299$ & $0.0130$ & $1{,}466$ & $342$ & $5.7\!\times\!10^{-80}$ \\
\addlinespace
\multicolumn{10}{l}{\textbf{Implicit}} \\
FLM-Audio vs Freeze-Omni & $0.3994$ & $0.8211$ & $11{,}937$ & $<10^{-300}$ & $0.0000$ & $0.0000$ & $87$ & $0$ & $1.000$ \\
FLM-Audio vs MiniCPM-o 4.5 & $0.3994$ & $0.9283$ & $11{,}481$ & $<10^{-300}$ & $0.0006$ & $0.4645$ & $1{,}804$ & $837$ & $1.2\!\times\!10^{-250}$ \\
FLM-Audio vs Moshi & $0.3994$ & $0.2705$ & $8{,}962$ & $1.1\!\times\!10^{-169}$ & $0.0006$ & $0.0146$ & $1{,}577$ & $24$ & $1.7\!\times\!10^{-4}$ \\
FLM-Audio vs Moshika-RL-Seamless & $0.3994$ & $0.4803$ & $9{,}784$ & $8.6\!\times\!10^{-61}$ & $0.0005$ & $0.0221$ & $1{,}898$ & $43$ & $5.6\!\times\!10^{-10}$ \\
FLM-Audio vs PersonaPlex & $0.3994$ & $0.0400$ & $8{,}047$ & $<10^{-300}$ & $0.0005$ & $0.0169$ & $1{,}892$ & $33$ & $4.4\!\times\!10^{-7}$ \\
FLM-Audio vs Realtime-Venus-Audio & $0.3994$ & $0.6087$ & $9{,}808$ & $<10^{-300}$ & $0.0006$ & $0.2327$ & $1{,}753$ & $409$ & $3.5\!\times\!10^{-119}$ \\
FLM-Audio vs Voila & $0.3994$ & $0.8427$ & $10{,}836$ & $<10^{-300}$ & $0.0006$ & $0.0166$ & $1{,}564$ & $27$ & $2.3\!\times\!10^{-5}$ \\
Freeze-Omni vs MiniCPM-o 4.5 & $0.8211$ & $0.9283$ & $4{,}992$ & $1.3\!\times\!10^{-214}$ & $0.0000$ & $0.4556$ & $90$ & $41$ & $5.1\!\times\!10^{-11}$ \\
Freeze-Omni vs Moshi & $0.8211$ & $0.2705$ & $13{,}061$ & $<10^{-300}$ & $0.0000$ & $0.0000$ & $79$ & $0$ & $1.000$ \\
Freeze-Omni vs Moshika-RL-Seamless & $0.8211$ & $0.4803$ & $10{,}207$ & $<10^{-300}$ & $0.0000$ & $0.0111$ & $90$ & $1$ & $1.000$ \\
Freeze-Omni vs PersonaPlex & $0.8211$ & $0.0400$ & $16{,}310$ & $<10^{-300}$ & $0.0000$ & $0.0222$ & $90$ & $2$ & $1.000$ \\
Freeze-Omni vs Realtime-Venus-Audio & $0.8211$ & $0.6087$ & $8{,}869$ & $<10^{-300}$ & $0.0000$ & $0.2099$ & $81$ & $17$ & $8.5\!\times\!10^{-4}$ \\
Freeze-Omni vs Voila & $0.8211$ & $0.8427$ & $6{,}041$ & $9.6\!\times\!10^{-7}$ & $0.0000$ & $0.0141$ & $71$ & $1$ & $1.000$ \\
MiniCPM-o 4.5 vs Moshi & $0.9283$ & $0.2705$ & $14{,}071$ & $<10^{-300}$ & $0.4690$ & $0.0168$ & $1{,}546$ & $713$ & $4.7\!\times\!10^{-197}$ \\
MiniCPM-o 4.5 vs Moshika-RL-Seamless & $0.9283$ & $0.4803$ & $10{,}487$ & $<10^{-300}$ & $0.4643$ & $0.0229$ & $1{,}876$ & $852$ & $1.1\!\times\!10^{-228}$ \\
MiniCPM-o 4.5 vs PersonaPlex & $0.9283$ & $0.0400$ & $18{,}148$ & $<10^{-300}$ & $0.4641$ & $0.0182$ & $1{,}868$ & $845$ & $2.4\!\times\!10^{-238}$ \\
MiniCPM-o 4.5 vs Realtime-Venus-Audio & $0.9283$ & $0.6087$ & $7{,}901$ & $<10^{-300}$ & $0.4653$ & $0.2364$ & $1{,}743$ & $629$ & $2.3\!\times\!10^{-59}$ \\
MiniCPM-o 4.5 vs Voila & $0.9283$ & $0.8427$ & $3{,}845$ & $2.6\!\times\!10^{-178}$ & $0.4708$ & $0.0175$ & $1{,}540$ & $718$ & $7.8\!\times\!10^{-193}$ \\
Moshi vs Moshika-RL-Seamless & $0.2705$ & $0.4803$ & $8{,}646$ & $<10^{-300}$ & $0.0159$ & $0.0239$ & $1{,}632$ & $55$ & $1.000$ \\
Moshi vs PersonaPlex & $0.2705$ & $0.0400$ & $5{,}755$ & $<10^{-300}$ & $0.0159$ & $0.0165$ & $1{,}633$ & $43$ & $1.000$ \\
Moshi vs Realtime-Venus-Audio & $0.2705$ & $0.6087$ & $10{,}682$ & $<10^{-300}$ & $0.0158$ & $0.2317$ & $1{,}515$ & $355$ & $7.1\!\times\!10^{-81}$ \\
Moshi vs Voila & $0.2705$ & $0.8427$ & $13{,}298$ & $<10^{-300}$ & $0.0154$ & $0.0147$ & $1{,}363$ & $39$ & $1.000$ \\
Moshika-RL-Seamless vs PersonaPlex & $0.4803$ & $0.0400$ & $9{,}619$ & $<10^{-300}$ & $0.0223$ & $0.0173$ & $1{,}969$ & $70$ & $1.000$ \\
Moshika-RL-Seamless vs Realtime-Venus-Audio & $0.4803$ & $0.6087$ & $9{,}654$ & $5.2\!\times\!10^{-156}$ & $0.0219$ & $0.2303$ & $1{,}828$ & $417$ & $5.4\!\times\!10^{-93}$ \\
Moshika-RL-Seamless vs Voila & $0.4803$ & $0.8427$ & $10{,}242$ & $<10^{-300}$ & $0.0203$ & $0.0166$ & $1{,}624$ & $52$ & $1.000$ \\
PersonaPlex vs Realtime-Venus-Audio & $0.0400$ & $0.6087$ & $12{,}085$ & $<10^{-300}$ & $0.0181$ & $0.2312$ & $1{,}821$ & $420$ & $1.4\!\times\!10^{-96}$ \\
PersonaPlex vs Voila & $0.0400$ & $0.8427$ & $16{,}525$ & $<10^{-300}$ & $0.0173$ & $0.0167$ & $1{,}619$ & $49$ & $1.000$ \\
Realtime-Venus-Audio vs Voila & $0.6087$ & $0.8427$ & $8{,}504$ & $<10^{-300}$ & $0.2372$ & $0.0179$ & $1{,}509$ & $359$ & $5.2\!\times\!10^{-82}$ \\
\addlinespace
\bottomrule
\end{tabular}
}
\end{table}

\end{document}